\documentclass[12pt]{article}
\usepackage{latexsym}
\usepackage{amssymb}
\catcode`\@=11
\newcount\hour
\newcount\minute
\newtoks\amorpm \hour=\time\divide\hour by 60\minute
=\time{\multiply\hour by 60 \global\advance\minute by-\hour}
\edef\standardtime{{\ifnum\hour<12 \global\amorpm={am}%
        \else\global\amorpm={pm}\advance\hour by-12 \fi
        \ifnum\hour=0 \hour=12 \fi
        \number\hour:\ifnum\minute<10
        0\fi\number\minute\the\amorpm}}
\edef\militarytime{\number\hour:\ifnum\minute<10
0\fi\number\minute}
\def\draftlabel#1{{\@bsphack\if@filesw {\let\thepage\relax
   \xdef\@gtempa{\write\@auxout{\string
      \newlabel{#1}{{\@currentlabel}{\thepage}}}}}\@gtempa
   \if@nobreak \ifvmode\nobreak\fi\fi\fi\@esphack}
        \gdef\@eqnlabel{#1}}
\def\@eqnlabel{}
\def\@vacuum{}
\def\marginnote#1{}
\def\draftmarginnote#1{\marginpar{\raggedright\scriptsize\tt#1}}
\def\draft{
        \pagestyle{plain}
        \overfullrule=2pt
        \oddsidemargin -.1truein
        \def\@oddhead{\sl \phantom{\today\quad\militarytime} \hfil
        \smash{\Large\sl DRAFT} \hfil \today\quad\militarytime}
        \let\@evenhead\@oddhead
        \let\label=\draftlabel
        \let\marginnote=\draftmarginnote
        \def\ps@empty{\let\@mkboth\@gobbletwo
        \def\@oddfoot{\hfil \smash{\Large\sl DRAFT} \hfil}
        \let\@evenfoot\@oddhead}
        \def\@eqnnum{(\theequation)\rlap{\kern\marginparsep\tt\@eqnlabel}%
        \global\let\@eqnlabel\@vacuum}  }

\renewcommand{\theequation}{\thesection.\arabic{equation}}
\renewcommand{\thefootnote}{\fnsymbol{footnote}}
\newcommand{\newsection}{    % Numeration of eqs. is automatic
\setcounter{equation}{0}\section}
\def\appendix#1{\addtocounter{section}{1}\setcounter{equation}{0}
\renewcommand{\thesection}{\Alph{section}}
\section*{Appendix \thesection\protect\indent \parbox[t]{11.15cm}{#1}}
\addcontentsline{toc}{section}{Appendix \thesection\ \ \ #1}}

\def\root{\sqrt{2}}

\def \bi{\bibitem}
\def \la {\label}
\def \ov {\over}

\def \b {\beta}
\def \Om {\Omega}

\def\qu{\frac{1}{4}}
\def\r2{\sqrt{2}}
\def \d {\partial}
\def \del{\partial}

\def\be{\begin{equation}}
\def\ee{\end{equation}}

\catcode`\@=11

\def\bea{\begin{eqnarray}}
\def\eea{\end{eqnarray}}
\def\beann{\begin{eqnarray*}}
\def\eeann{\end{eqnarray*}}
\def\beq{\begin{equation}}
\def\eeq{\end{equation}}
\def\ba{\begin{array}}
\def\ea{\end{array}}
\def\ben{\begin{enumerate}}
\def\een{\end{enumerate}}
 \def \l {\lambda}

 \def \la {\label}
 \def\be{\begin{equation}}
\def\ee{\end{equation}}

\def \la {\label}

\def \r {\rho}

\font\mybb=msbm10 at 11pt

\def\bb#1{\hbox{\mybb#1}}

\def\bR {\bb{R}}

\def\e  {\epsilon}

\def \ov {\over}
\def \ha { { 1\ov 2}}

\def \del { \partial}

\def \ee {\epsilon}

\def \g {\gamma}
\def \bi{\bibitem}
\def\a{\alpha }

\def \r {\rho}
\def \d {\delta}
\def \G {\Gamma}
\def \l {\lambda}

\def \g {\gamma}

\def \b {\beta}

\def\be{\begin{equation}}
\def\ee{\end{equation}}

\def \bi {\bibitem}
\def \la{\label}

\def \fff {\vrule width 0.5pt height 6pt depth 1pt}
\def \pp { {=\hskip -4.75pt {\fff}\hskip 3.75pt}}

\begin{document}
%\draft
\date{November 2002}
%%%%%%%%%%%%%%%%%%%%%%%%%%%%%%%%%%%%%%%%%%%%%%%%%%%%%%%%%%
\begin{titlepage}
\begin{center}
%\today
\hfill hep-th/0602250\\
\hfill KUL-TF-06/04\\

\vspace{1.5cm}
{\Large \bf  Geometry of type II  common sector N=2 backgrounds}\\[.2cm]

\vspace{1.5cm}
 {\large U.~Gran$^1$,  P.~Lohrmann$^2$ and  G.~Papadopoulos$^2$ }

 \vspace{0.5cm}
${}^1$ Institute for Theoretical Physics, K.U. Leuven\\
Celestijnenlaan 200D\\
B-3001 Leuven, Belgium\\
\vspace{0.5cm}

${}^2$ Department of Mathematics\\
King's College London\\
Strand\\
London WC2R 2LS
\end{center}

\vskip 1.5 cm
\begin{abstract}

We describe the geometry of all type II common sector backgrounds
with two supersymmetries. In particular, we determine the spacetime
geometry of those supersymmetric backgrounds for which each copy of
the Killing spinor equations
 admits a Killing spinor. The stability subgroups
of both Killing spinors are $Spin(7)\ltimes \bR^8$, $SU(4)\ltimes \bR^8$ and $G_2$ for IIB backgrounds, and
$Spin(7)$, $SU(4)$ and $G_2\ltimes \bR^8$  for IIA backgrounds. We show that the spacetime of backgrounds
with spinors that have stability subgroup $K\ltimes \bR^8$ is a pp-wave propagating in an eight-dimensional manifold
 with a $K$-structure. The spacetime of  backgrounds with $K$-invariant Killing spinors
is a fibre bundle with fibre spanned by the orbits of two commuting
null Killing vector fields and base space an eight-dimensional
manifold which admits a $K$-structure. Type II T-duality
interchanges the backgrounds with $K$- and $K\ltimes\bR^8$-invariant
Killing spinors. We show that the geometries of the base space of
the fibre bundle and the corresponding space in which the pp-wave
propagates are the same. The
conformal symmetry of the world-sheet action of
 type II strings propagating in these $N=2$ backgrounds can always be fixed in the light-cone gauge.

\end{abstract}
\end{titlepage}
%%%%%%%%%%%%%%%%%%%%%%%%%%%%%%%%%%%%%%%%%%%%%%%
\newpage
\setcounter{page}{1}
\renewcommand{\thefootnote}{\arabic{footnote}}
\setcounter{footnote}{0}

\setcounter{section}{0}
\setcounter{subsection}{0}
%%%%%%%%%%%%%%%%%%%%%%%%%%%%%%%%%%%
\newsection{Introduction}

The type II common sector is a consistent truncation of type II supergravities. The bosonic
fields are the metric $g$, dilaton $\Phi$
and three-form field strength $H$, $dH=0$. The Killing spinor equations in the string frame are
\bea
&&\hat\nabla\hat\epsilon=0~,~~~(d\Phi-{1\over 2}H)\hat\epsilon=0~,
\cr
&&\check\nabla\check\epsilon=0~,~~~(d\Phi+{1\over 2}H)\check\epsilon=0~,
\la{kse}
\eea
where   the Killing spinors $\hat\epsilon$ and $\check\epsilon$
 are Majorana-Weyl of the same (IIB) or opposite chirality (IIA), and $\hat\nabla=\nabla+{1\over2} H$
 and $\check\nabla=\nabla-{1\over2} H$
are metric connections with torsion\footnote{In the literature,
these connections are denoted with $\nabla^+$ and $\nabla^-$,
respectively. We have introduced a different notation because later
we use $-$ and $+$ labels to denote light-cone directions.} given by
the NS$\otimes$NS three-form $H$. The gravitino and dilatino Killing
spinor equations   are  two copies of those of the heterotic string
differing by the sign of $H$. Let $(\hat N, \check N)$ be the number
of Killing spinors  and $(\hat G, \check G)$ be their stability
subgroups in $Spin(9,1)$ for each copy, respectively.  The total
number of Killing spinors of a background is $N=\hat N+\check N$. If
$\check N=0$, then the Killing spinor equations reduce to those of
the heterotic string. Consequently, the associated supersymmetric
backgrounds  are those of the heterotic string and a systematic
investigation of their  geometries  can be found in \cite{ugplgp}
using the spinorial geometry approach of \cite{grangp}. There is
extensive previous work on the geometry of heterotic and type II
common sector supersymmetric backgrounds, see
e.g.~\cite{strominger}-\cite{lustb}. Similarly, if $\hat N=0$, the
geometries of the common sector backgrounds can also be recovered
from those of the heterotic string. However, if both sectors admit a
non-trivial Killing spinor, $\hat N, \check N\not=0$, new geometries
arise. The conditions that  each copy of Killing spinor equations
imposes on the geometry are those  of, or can be constructed from,
the associated heterotic string backgrounds with $\hat N$ and
$\check N$ Killing spinors and with stability subgroups $\hat G$ and
$\check G$, respectively. Furthermore, the geometry of the spacetime
depends on the conditions that arise from both copies of the Killing
spinor equations and in particular on the stability subgroup,
$G=\hat G\cap \check G$, of all Killing spinors\footnote{The use of
the stability subgroups of the spinors in the context of
supersymmetric solutions has been suggested in \cite{jose}.}. In
turn, $G$ depends on the way that $\hat G$ and $\check G$ are
embedded in $Spin(9,1)$ up to a $Spin(9,1)$-conjugation. There are
many ways to embed $\hat G$ and $\check G$ in $Spin(9,1)$
 which
lead to inequivalent geometries for the spacetime.
 In \cite{ugplgp} it was argued that  all these new geometries
appear because, unlike
the case of the heterotic string, the gauge group $Spin(9,1)$ of the common sector Killing spinor equations  is
a proper subgroup of the holonomy group $Spin(9,1)\times Spin(9,1)$
 of the supercovariant connection of the gravitino Killing spinor equations.

All $N=1$ supersymmetric common sector backgrounds\footnote{It suffices to know
 the geometry of $\hat N=1$, $\check N=0$ backgrounds. This is  because the geometry of $\hat N=0$, $\check N=1$, backgrounds
 can be easily be determined from that of  $\hat N=1$, $\check N=0$ backgrounds, e.g.~in IIB common
 sector one has to set $H\rightarrow -H$.}
 are embeddings
of $N=1$ supersymmetric heterotic backgrounds and therefore their
geometry has already been described in \cite{ugplgp}. As we have
already mentioned, the geometry of $N=2$ backgrounds with either
$\hat N=0$ or $\check N=0$ are also embeddings of heterotic string
backgrounds with two Killing spinors. In this paper, we shall
examine the geometry of type II common sector $N=2$ backgrounds with
$\hat N=\check N=1$. It has been shown in \cite{ugplgp} that there
are three classes of such type IIB backgrounds characterized by the
spinor stability subgroups $G=Spin(7)\ltimes \bR^8$, $SU(4)\ltimes
\bR^8$ and $G_2$. We shall  show that there are also three classes
of IIA backgrounds characterized by the spinor stability subgroups
$G=Spin(7)$, $SU(4)$ and $G_2\ltimes \bR^8$. It is clear that there
are two types of stability subgroups $K$ and $K\ltimes \bR^8$, where
$K$ is a compact group, and the type II N=2 backgrounds come in
pairs. If a IIA background has stability subgroup $G=K$, then the
corresponding IIB background has stability subgroup $G=K\ltimes
\bR^8$ and vice-versa. All the backgrounds with stability subgroup
$G=K\ltimes \bR^8$ admit a  null $\nabla$-parallel vector field and
therefore the spacetime is a pp-wave propagating in an
eight-dimensional space $B$ equipped with a $K$-structure.
Equivalently, the spacetime can be thought of as a Lorentzian
deformation family of $B$. In particular, the metric and three-form
are
\bea
&&ds^2=2 dv (du+ V dv+n)+ \delta_{ij} e^i e^j~,
\cr
&&H={1\over2} H^{{\mathfrak k}}_{-ij} e^-\wedge e^i\wedge e^j+{1\over2} H^{{\mathfrak k}^\perp}_{-ij} e^-\wedge e^i\wedge e^j
+{1\over 3!} H_{ijk} e^i\wedge e^j\wedge e^k~,
\cr
&&\Phi=\Phi(v,y)~,
\la{nullback}
\eea
where the parallel vector field is $X=\partial/\partial u$ and the
deformed manifold $B$ is defined by $u,v={\rm const}$. Let
$\mathfrak{k}\subset \Lambda^2(\bR^8)$ be the Lie algebra of $K$ and
$\Lambda^2(\bR^8)=\mathfrak{k}\oplus \mathfrak{k}^\perp$. The
Killing spinor equations specify all the components of $H$ in terms
of the geometry apart from $H^{{\mathfrak k}}_{-ij}$ which remains
undetermined.
 The geometry of the eight-dimensional deformed manifold is constrained. In particular for $K=Spin(7)$, $B$ is a $Spin(7)$ manifold,
 ${\rm hol}(\tilde\nabla)\subseteq Spin(7)$,
 and $H_{ijk}=0$. For $K=SU(4)$,  $B$ is an almost hermitian manifold with an $SU(4)$ structure, the canonical bundle of $B$ admits a trivialization
 and the associated (4,0)-form is parallel with respect to the Levi-Civita connection. The classes $W_1$ and $W_4$ are
 related to the trivialization of the
 canonical bundle, and  $W_2$ is related to the $W_3$ class. For $K=G_2$, $B$ admits a vector field $Z$ which is rotation free but not Killing.
 There are $2^{10}$ $G_2$-structures on an eight-dimensional manifold and we specify the one that is associated with $N=2$ supersymmetry.
 The components $H_{ijk}$ of the torsion are determined in terms of $Z$ and its derivatives. In addition, we analyze the integrability
 conditions of the Killing spinor equation and we show that all the field equations are satisfied provided that one imposes
 the Bianchi identity of $H$, $dH=0$, and the $E_{--}=0$ component of the Einstein equations.

 The backgrounds with stability subgroups $K$ are (locally) fibre bundles of rank two over an eight-dimensional manifold $B$.
 The fibre directions are spanned by the orbits of two commuting null $\hat\nabla$- and $\check \nabla$-parallel
 vector fields $X,Y$. The metric and torsion can be written as
 \bea
 &&ds^2=2 f^4 (dv+m) (du+n)+\delta_{ij} e^i e^j~,
 \cr
&& H= d(e^-\wedge e^+)+{1\over3 !}H_{ijk} e^i\wedge e^j\wedge e^k~,
 \cr
&& \Phi=\Phi(y)~,
\la{timeback}
 \eea
where $X=\partial/\partial u$ and $Y=\partial/\partial v$. All the
components of the torsion $H$ are determined in terms
of the geometry of spacetime. In particular, if $G=Spin(7)$,
$H_{ijk}=0$. If the stability subgroup of the spinors in $K$, we find that the geometry of the base space $B$ is exactly the same
as that of the deformed manifold $B$ for the backgrounds with $K\ltimes\bR^8$-invariant spinors
described above. As in the previous case, we analyze the
integrability
 conditions of the Killing spinor equations and show that  all the field equations are satisfied provided that one imposes
 the Bianchi identity of $H$, $dH=0$ and $LH_{-+}=0$.

 We also investigate the dynamics of fundamental string probes in the above supersymmetric backgrounds.
 As an application of the geometry of $\hat N=\check N=1$ backgrounds, we show that
  conformal symmetry of the world-sheet action of strings propagating in $\hat N, \check N\geq 1$ supersymmetric backgrounds
  can always be fixed in the light-cone gauge. In addition, we  give the bosonic part of the light-cone world-sheet actions.
  Then we investigate the relation between spacetime supersymmetry and the world-sheet chiral $W$-type of currents of \cite{howegpw}
  for string probes.

This paper has been organized as follows: In section two, we give
the representatives of the Killing spinors and their stability
subgroups in $Spin(9,1)$, and describe some of the properties of the
Killing spinor form bilinears. In section three, we solve the
Killing spinor equations of $\hat N=\check N=1$ type IIB
supersymmetric backgrounds and we investigate the associated
geometries. In section four, we solve the Killing spinor equations
of $\hat N=\check N=1$ type IIA
 supersymmetric backgrounds and
we investigate the associated geometries. In section five, we give
the light-cone action of strings in type II backgrounds with $\hat
N,\check N\geq 1$ supersymmetry and in section six, we present our
conclusions. In appendix A, we summarize the type II Killing spinor
equations, their integrability conditions and the field equations of
the theory. In appendix B, we evaluate the type IIA Killing spinor
equations on a basis in the space of negative chirality spinors.
This together with the results of \cite{ugplgp} provides all the
data one needs for the systematics of type II common sector. In
appendix C, we summarize some of the properties of eight-dimensional
manifolds with $SU(4)$ and $G_2$ geometries.

\newsection{Spinors, holonomy and forms}

\subsection{Holonomy, gauge symmetry and parallel spinors}

The spinor bundle of the IIB common sector is $S^+\oplus S^+$  and
the Killing spinor equations are two copies of those of the
heterotic string differing by the sign of the NS$\otimes$ NS
three-form field strength, see (\ref{kse}). In particular, the
gravitino Killing spinor equation is a parallel transport equation
for  the $Spin(9,1)\times Spin(9,1)$ connection $\hat \nabla\oplus
\check \nabla$. The gauge group that preserves the Killing spinor
equations is  $Spin(9,1)$. This is in contrast to the heterotic case
where the holonomy group of the connection of the gravitino Killing
spinor equation coincides with the gauge group. Since the gauge
group of the common sector is the same as that of the heterotic
string but the dimension of the space of spinors is twice as large,
there are many more cases to investigate than those examined for the
heterotic string in \cite{ugplgp}.

The systematics of type IIB common sector Killing spinor equations
can be read off from those of the heterotic string in \cite{ugplgp}.
This is because the representation of the Killing spinors is two
copies of that of the heterotic string, i.e.~$\Delta_{{\bf
16}}^+\oplus \Delta_{{\bf 16}}^+$. In particular, the linear system
of the first copy of the Killing spinor equations is exactly the
same as that of the heterotic string. The linear system of the
second copy can be read off from that of the heterotic string by
setting $H\rightarrow -H$.

As a consequence, the conditions  that each copy of the Killing spinor equations imposes on the geometry
of spacetime can be found from those of the heterotic string in \cite{ugplgp}. In particular,
${\rm hol}(\hat\nabla)\subseteq \hat G$ and ${\rm hol}(\check\nabla)\subseteq \check G$, where
$\hat G$ and $\check G$ are the stability subgroups of the Killing spinors of each copy in $Spin(9,1)$.
However, the geometry of spacetime depends on the conditions of both copies and in particular
of the stability subgroup $G$ of all Killing spinors in $Spin(9,1)$. In turn, $G$ depends on the embedding
of $\hat G$ and $\check G$ in $Spin(9,1)$ up to a conjugation with a $Spin(9,1)$ gauge transformation.

To illustrate the above analysis, consider the $N=2$ IIB common sector backgrounds with $\hat N=\check N=1$.
It has been shown in \cite{ugplgp} that the Killing spinors can be chosen as
\bea
\hat\epsilon=f(1+e_{1234})~,~~~ \check\epsilon= g_1 (1+e_{1234})+ i
g_2 (1-e_{1234})+g_3 (e_{15}+ e_{2345})~, \la{iibks}
\eea
where $f, g_1, g_2$ and $g_3$ are spacetime functions. Both the
spinor $\hat\epsilon$ and $\check\epsilon$ are representatives of
the $Spin(7)\ltimes \bR^8$ orbit of $Spin(9,1)$ in the Majorana-Weyl
representation  $\Delta_{\bf 16}^+$.  Therefore $\hat G=\check
G=Spin(7)\ltimes \bR^8$.
 The second spinor $\check \epsilon$ has been constructed
by decomposing $\Delta_{\bf 16}^+$ under  $Spin(7)\ltimes \bR^8$
representations and taking suitable representatives of the orbits.
 The stability subgroup of both spinors depends on the coefficients
$g_1, g_2$ and $g_3$. Note that if $g_3\not=0$, one can set
$g_1=g_2=0$ by acting with an $\bR^8\subset Spin(7)\ltimes\bR^8$
transformation which stabilizes $\hat\epsilon$, see \cite{ugplgp}
for details. In  table 1,  we summarize the
 stability subgroups of the Killing spinors of $N=2$ IIB common sector backgrounds.

 \begin{table}[h]
 \begin{center}
\begin{tabular}{|c|c|c|c|c|}\hline
$ {\rm IIB}, N=2$&$\mathrm{\hat G }$ & $\mathrm{\check G }$& $\mathrm{G }$&{\rm Spinors}
 \\ \hline
 &$SU(4)\ltimes \bR^8 $  &- &$SU(4)\ltimes \bR^8 $& $-$ \\
 &$G_2$ &- &$G_2$ &$-$\\
&$Spin(7)\ltimes \bR^8$&$Spin(7)\ltimes \bR^8$& $Spin(7)\ltimes \bR^8$&$g_1\not=0, g_2=g_3=0$ \\
&$Spin(7)\ltimes \bR^8$&$Spin(7)\ltimes \bR^8$&$SU(4)\ltimes \bR^8$& $g_1,g_2\not=0, g_3=0$\ \\
&$Spin(7)\ltimes \bR^8$&$Spin(7)\ltimes \bR^8$&$G_2$&$g_1=g_2=0$, $g_3\not=0$
\\ \hline
\end{tabular}
\end{center}
\vskip 0.2cm {\small Table 1:  There are five classes of IIB common
sector backgrounds with two supersymmetries up to $Spin(9,1)$ gauge transformations. These are denoted with
the stability subgroups $\hat G$, $\check G$ and $G$ of the Killing spinors.
In all cases ${\rm hol}(\hat\nabla)\subseteq \hat G$ and ${\rm hol}(\check\nabla)\subseteq \check G$. The entries $-$
denote the  cases for which the sector associated with the
$\check \nabla$ connection  does not admit Killing spinors. The last column gives the restrictions
on the parameters of the second Killing spinor in (\ref{iibks}).}

\end{table}

The Killing spinor equations of the type IIA common sector are
somewhat different from those of the type  IIB common sector. This is
because the type IIA spinor bundle is $S^+\oplus S^-$ and therefore
is not just two copies of that of the heterotic string.
Nevertheless, the systematics of the IIA common sector Killing
spinor equations can again be read off from those of the heterotic
string in \cite{ugplgp}. In particular, the linear system for the
first copy is identical to that of the heterotic string. The linear
system of the second copy of the Killing spinor equations has also some similarities with that of
the heterotic string and is given in appendix B.

The analysis of the stability subgroups of spinors in the IIA common sector
 and their relation to the geometry is similar to the one we have presented
for the IIB common sector. However there are some differences due to
the different representations of the Killing spinors. This can be
seen in the $N=2$ backgrounds with $\hat N=\check N=1$. One can
again choose $\hat\epsilon=f (1+e_{1234})$. It remains to find
representatives of the second spinor in $\Delta_{{\bf 16}}^-$ up to
$Spin(7)\ltimes\bR^8$ transformations which is the stability
subgroup of $\hat\epsilon$ in $Spin(9,1)$. One can show that under
$Spin(7)$, $\Delta_{\bf 16}^-$ decomposes as
$\bR\oplus\bR^7\oplus\Delta^+_{\bf 8}$. The spinor that represents
the singlet in the decomposition is proportional to $e_5+e_{12345}$.
The stability subgroup of $1+e_{1234}$ and $e_5+e_{12345}$ is
$Spin(7)$ rather than $Spin(7)\ltimes\bR^8$ that appears in the IIB
case. The second spinor can be written as $\check\epsilon=g_1
(e_5+e_{12345})+\epsilon_2+\epsilon_3$, where $\epsilon_2$ and
$\epsilon_3$ lie in the seven and eight-dimensional representations
respectively. If the component $\epsilon_3$ vanishes, then $Spin(7)$
acts transitively on the sphere in $\bR^7$ and so the representative
can be chosen to lie in any direction. In particular, one can choose
as the second spinor $\check\epsilon=g_1 (e_5+e_{12345})+ig_2
(e_5-e_{12345})$. The stability subgroup of both spinors is $SU(4)$.
Next suppose that $\epsilon_3$ does not vanish. To choose
$\epsilon_3$ observe that $Spin(7)$ acts transitively on the sphere
in $\Delta^+_{\bf 8}$ with stability subgroup $G_2$. In turn $G_2$
acts transitively in $\bR^7$ with stability subgroup $SU(3)$. To
summarize, we have found
 that
one can always choose the two Killing spinors as
\bea
\hat\epsilon=f (1+e_{1234})~,~~~\check\epsilon=g_1(e_5+e_{12345})+ i g_2 (e_5- e_{12345})+ g_3 (e_1+e_{234})~.
\la{iiaks}
\eea
There is one type of orbit of $Spin(9,1)$ in $\Delta_{{\bf 16}}^-$
with stability subgroup $Spin(7)\ltimes \bR^8$, thus the stability
subgroup of either $\hat\epsilon$ or $\check \epsilon$ is
$Spin(7)\ltimes\bR^8$, i.e.~$\hat G=Spin(7)\ltimes\bR^8$ and $\check
G=Spin(7)\ltimes\bR^8$. To find the stability subgroups of both
$\hat\epsilon$ and $\check\epsilon$, observe that if  $g_1\not=0$,
the element
\bea
g=e^{{1\over \sqrt{2} (g_1^2+g_2^2)} (g_1 g_3\Gamma^1 \Gamma^--g_2 g_3 \Gamma^6 \Gamma^-)}
\eea
of $Spin(9,1)$ leaves invariant $\hat\epsilon$ and transforms
the spinor $g_1(e_5+e_{12345})+ i g_2 (e_5- e_{12345})$ to $\check \epsilon$ in (\ref{iiaks}),
and similarly for $g_2\not=0$. Therefore,
 if either $g_1\not=0$ or $g_2\not=0$, one can always choose $g_3=0$. This is  in analogy with a similar
result in type IIB \cite{ugplgp}. The stability subgroups of
$\hat\epsilon$ and $\check\epsilon$ in the IIA common sector are
summarized in table 2.

\begin{table}[h]
 \begin{center}
\begin{tabular}{|c|c|c|c|c|}\hline
${\rm IIA}, N=2$&$\mathrm{\hat G }$ & $\mathrm{\check G }$& $\mathrm{G }$& ${\rm Spinors}$
 \\ \hline
 &$SU(4)\ltimes \bR^8 $  &- &$SU(4)\ltimes \bR^8 $& - \\
 &$G_2$ &- &$G_2$& - \\
&$Spin(7)\ltimes \bR^8$&$Spin(7)\ltimes \bR^8$& $Spin(7)$&$g_1\not=0, g_2=g_3=0$ \\
&$Spin(7)\ltimes \bR^8$&$Spin(7)\ltimes \bR^8$&$SU(4)$& $g_1, g_2\not=0, g_3=0$ \\
&$Spin(7)\ltimes \bR^8$&$Spin(7)\ltimes \bR^8$&$G_2\ltimes\bR^8$& $g_1=g_2=0, g_3\not=0$
%\\&$Spin(7)\ltimes \bR^8$&$Spin(7)\ltimes \bR^8$&$SU(3)$&$g_1, g_2, g_3\not=0$
\\
\hline
\end{tabular}
\end{center}
\vskip 0.2cm {\small Table 2:  There are five classes of IIA common
sector backgrounds with two supersymmetries. These are denoted with
the stability subgroups $\hat G$, $\check G$ and $G$ of the Killing spinors.
In all cases ${\rm hol}(\hat\nabla)\subseteq \hat G$ and  ${\rm hol}(\check\nabla)\subseteq \check G$.
The entries $-$
denote the  cases for which the copy associated with the
$\check \nabla$ connection  does not admit Killing spinors.  The last column gives the restrictions
on the parameters of the second Killing spinor in (\ref{iiaks}).}

\end{table}

 The stability subgroups of the spinors in the type IIA and type IIB common sector backgrounds with two supersymmetries
 are different in all $\hat N, \check N \not=0$ cases. However, they are related by the interchange $K\leftrightarrow K\ltimes\bR^8$, where
 $K=Spin(7), SU(4)$ and $G_2$. We shall argue later that this is due to the type II T-duality.

\subsection{Spacetime form bilinears}

There are three kinds of spacetime form spinor bilinears that one can construct for the type II common sector. One kind is
the form bilinears that are constructed from  $\hat \nabla$-parallel spinors. Another kind are those
that are constructed from $\check \nabla$-parallel spinors and the third kind are those which are constructed from one $\hat \nabla$- and
one $\check \nabla$-parallel spinor.  We denote these bilinears with $\hat\alpha$, $\check \alpha$ and $\alpha$, respectively.
It is clear that
\bea
\hat \nabla_A\hat\alpha=0~,~~~~\check \nabla_A\check\alpha=0~.
\eea
The bilinears $\alpha$ are not apparently parallel with neither $\hat \nabla$ nor $\check \nabla$ connections. Instead, one finds that
\bea
&&\nabla_A\a={1\over k!}B( \nabla_A\hat \psi, \Gamma_{B_1\dots B_k}\check \eta)e^{B_1}\wedge\dots\wedge e^{B_k}
+ {1\over k!}B(\hat \psi, \Gamma_{B_1\dots B_k}\nabla_A\check \eta)e^{B_1}\wedge\dots\wedge e^{B_k}
\cr
&&~~~~~=-{1\over8\cdot k!} H_{AC_1C_2} B(\hat \psi, \{\Gamma^{C_1C_2},\Gamma_{B_1\dots B_k}\}\check \eta)e^{B_1}\wedge\dots\wedge e^{B_k}~.
\eea
The computation of the form bilinears can be  done as
for the heterotic string \cite{ugplgp}.

A special class of spinor bilinears are the one-forms. If
$\hat\kappa_X$ and $\check\kappa_Y$ are parallel one-forms with
respect to the connections $\hat \nabla$ and $\check \nabla$,
respectively, then the associated vector fields $\hat X$ and $\check
Y$ are Killing
\bea
{\cal L}_{\hat X} g=0~,~~~~~{\cal L}_{\check Y} g=0~,
\eea
 and
\bea
d\hat\kappa_X=i_{\hat X} H~,~~~~d\check\kappa_Y=-i_{\check Y} H~.
\eea
Since $dH=0$, clearly ${\cal L}_{\hat X} H={\cal L}_{\check Y} H=0$.

As in the case of the heterotic string,  the commutators of
$\hat\nabla$- and $\check\nabla$-parallel vector fields are
determined in terms of $H$. In particular, one finds that
\bea
&&[\hat X,\hat Y]=i_{\hat X} i_{\hat Y} H^A e_A~,~~~[\check X,\check Y]
=-i_{\check X} i_{\check Y} H^A e_A~,
\cr
&&[\hat X, \check Y]=0~.
\eea
Note that the commutator of $\hat \nabla$-parallel   with  $\check
\nabla$-parallel vector fields
 vanishes. This property is widely applicable because in all $N\geq 2$, $\hat N, \check N\geq 1$, backgrounds there is
at least one $\hat \nabla$-parallel and one $\check \nabla$-parallel
vector field. However, these vector fields are not always linearly
independent.  In addition if $\hat X, \hat Y$ and $\check X, \check
Y$ are $\hat\nabla$- and $\check \nabla$-parallel,
 then $[\hat X, \hat Y]$ and $[\check X, \check Y]$ are also $\hat\nabla$- and $\check \nabla$-parallel, respectively
 \cite{ugplgp}.

Suppose that $\hat\alpha$ and $\check \alpha$ are $k$-forms, and
that $\hat X$ and $\check X$ are vector fields. Then, one can show
using $\hat\nabla \hat\a=\hat\nabla \hat X=\check \nabla \check
\a=\check \nabla \check X=0$ that
\bea
&&({\cal L}_{\hat X}\hat\alpha)_{A_1\dots A_k}=k (-1)^k (i_{\hat X}H)^B{}_{[A_1} \hat\alpha_{A_2\dots A_k]B}~,
\cr
&&({\cal L}_{\check X}\check\alpha)_{A_1\dots A_k}=-k (-1)^k (i_{\check X}H)^B{}_{[A_1} \check\alpha_{A_2\dots A_k]B}~,
\cr
&&{\cal L}_{\hat X}\check\alpha={\cal L}_{\check X}\hat\alpha=0~.
\la{calxa}
\eea
Therefore ${\cal L}_{\hat X}\hat\alpha={\cal L}_{\check X}\check\alpha=0$, iff the rotations of $\hat X$
and $\check X$,   $i_{\hat X}H$
and $-i_{\check X}H$, leave invariant the forms $\hat\a$ and $\check\a$, respectively.

\newsection{IIB $N=2$ Backgrounds}

\subsection{Backgrounds with $Spin(7)\ltimes \bR^8$-invariant Killing spinors}

\subsubsection{Supersymmetry conditions}

The conditions that the Killing spinor equations impose on the
geometry of spacetime can be easily read off from those of the
heterotic string for backgrounds with one supersymmetry. As we have
explained, the $Spin(7)\ltimes\bR^8$-invariant Killing spinors   can
be chosen as
\bea
\hat\epsilon= f(1+e_{1234})~,~~~\check\epsilon= g (1+e_{1234})~.
\eea
The conditions that are implied by the spinor $\hat\epsilon$ are
exactly those of the $N=1$ heterotic string backgrounds found in
\cite{ugplgp}. The conditions that are implied by the spinor
$\check\epsilon$ can be read off from those of the $N=1$ heterotic
string backgrounds after substituting $H\rightarrow -H$. Because of
this, we shall not elaborate on the derivation of the linear system
associated with the Killing spinor equations. The relation between
linear systems and Killing spinor equations is explained in
\cite{grangpb}. The linear system can be solved to give
\bea
&&g=f~,~~~2\partial_A\log f+\Omega_{A,-+}=0~,~~~\Omega_{A,\a}{}^\a=0
\cr
&&\Omega_{A,\bar\a\bar\b}-{1\over2} \Omega_{A, \g\d} \epsilon^{\g\d}{}_{\bar\a\bar \b}=0~,~~~\Omega_{A,+\a}=0~,
\cr
&&\partial_\a\Phi=\partial_+\Phi=0~,~~~H_{ijk}=0~, ~~~H_{+AB}=0~,
\cr
&&H_{-\a}{}^\a=0~,~~~H_{-\bar\a\bar\b}-{1\over2} H_{-\g\d} \epsilon^{\g\d}{}_{\bar\a\bar \b}=0~,
\eea
where we have used that $f,g$ are defined up to an overall constant
scale. The last three conditions above imply that the only
non-vanishing components of the flux $H$ are $H_{-ij}$ and they take
values in $\mathfrak{spin}(7)\subset
\mathfrak{so}(8)=\Lambda^2(\bR^8)$. Using a spin gauge
transformation in the direction $\Gamma_{-+}$, one can  choose the
gauge  $f=1$, which implies $g=1$ as can be seen from the above
equations.
 In this gauge,
the Levi-Civita connection satisfies
\bea
\Omega_{A,+B}=0~. \la{apb}
\eea
This simplifies the investigation of the geometry. The geometry of
spacetime is independent of the choice of gauge.

\subsubsection{The geometry of spacetime}

The linearly independent spacetime form bilinears constructed from
the Killing spinors $\hat\epsilon= f(1+e_{1234})=\check\epsilon$,
after an appropriate normalization, are
\bea
\hat\kappa=\kappa(\hat\epsilon, \hat\epsilon)=f^2 (e^0- e^5)~,~~~\hat\tau=\tau(\hat\epsilon, \hat\epsilon)=f^2 (e^0-e^5)\wedge \phi~,
\la{bieone}
\eea
where
\bea
\phi={\rm Re} \chi-{1\over2} \omega\wedge \omega~,
\eea
is a $Spin(7)$-invariant form, $\chi=(e^1+i e^6)\wedge (e^2+i e^7)\wedge (e^3+i e^8)\wedge (e^4+i e^9)$ and
$\omega=-(e^1\wedge e^6+e^2\wedge e^7+e^3\wedge e^8+e^4\wedge e^9)$. Clearly $\hat\alpha=\check\alpha$, where
$\hat\alpha$ and $\check\alpha$ denote collectively all the bilinears constructed from $(\hat\epsilon, \hat\epsilon)$ and
$(\check\epsilon, \check\epsilon)$,
respectively.

As we have already explained,
$\hat\nabla\hat\alpha=\check\nabla\check\alpha=0$. Furthermore, in
the gauge $f=g=1$, $\hat\alpha=\check\alpha=\alpha$. This implies
that $\alpha$ is also parallel with respect to the Levi-Civita
connection. In particular, one finds that
\bea
\nabla_A \kappa_B=0~,~~~~\nabla_A \phi_{ijkl}=0~,
\eea
where $i,j=1,2,3,4,6,7,8,9$.
The former condition can also be easily seen from (\ref{apb}).
 Therefore the holonomy of the Levi-Civita connection, $\nabla$, is contained in
$Spin(7)\ltimes \bR^8$, ${\rm hol}(\nabla)\subseteq Spin(7)\ltimes \bR^8$.

A consequence of this is that the vector field $X$ associated with
$\kappa=e^-$ is rotation free. If one adapts coordinates along $X$,
$X=\partial/\partial u$, and uses that $X$ is rotation free, then
$e^-=dv$ and
 the metric, flux and dilaton
can be written as
\bea
&&ds^2=2 e^- e^++\delta_{ij} e^i e^j= 2 dv (du +V dv+ n_I dy^I)+ g_{IJ} dy^I dy^J~,
\cr
&&H={1\over2} H^{\mathfrak{spin}(7)}_{-ij} e^-\wedge e^i\wedge e^j~,~~~~
\cr
&&\Phi=\Phi(v)~,
\eea
where all the components of the  fields  depend on $v, y$ and
as is indicated $H_{-ij}$ takes values in $\mathfrak{spin}(7)$.

The spacetime is a pp-wave propagating in an eight-dimensional
$Spin(7)$ manifold $B$ given by $u,v={\rm const}$. The holonomy of
the Levi-Civita connection of $B$, $\tilde\nabla$, is contained in
$Spin(7)$, ${\rm hol}(\tilde\nabla)\subseteq Spin(7)$.
Alternatively, the spacetime can be thought of as a
two-parameter\footnote{The geometry of $B$ depends non-trivially
only on $v$.} Lorentzian family of eight-dimensional $Spin(7)$
holonomy manifolds $B$.

\subsubsection{Field equations}

The integrability conditions of the Killing spinor equations have
been given in appendix A. As is well-known, these can be used to
find which field equations are implied as a consequence of the
Killing spinor equations. A straightforward calculation using the
equations in appendix A and the results in appendix B
 reveals that if  one imposes
\bea
dH=0~,~~~E_{--}=0~,
\eea
then all the rest of the field equations are implied. This is the
case for all the $\hat N=\check N=1$ supersymmetric common sector
backgrounds which admit $K\ltimes\bR^8$-invariant Killing spinors.
As the proof is very similar for the other cases we shall not repeat
the analysis in each case.

\subsection{Backgrounds with $SU(4)\ltimes \bR^8$-invariant Killing spinors}
\subsubsection{Supersymmetry conditions}

The $SU(4)\ltimes \bR^8$-invariant Killing spinors can be chosen as
\bea
\hat\epsilon= f(1+e_{1234})~,~~~\check\epsilon= (g_1+i g_2) 1+ (g_1-ig_2) e_{1234}~.
\eea
The linear system  associated with the above Killing spinors can be
easily derived from the heterotic supergravity results of
\cite{ugplgp} so we shall not give further details. The solution of
the linear system can be written as
\bea
&&\Omega_{A,+B}=0~,~~~H_{+AB}=0~,~~~\hat\Omega_{A,\a}{}^\a=0,
\cr
&&\partial_A (g_1+ig_2)+\Omega_{A,\a}{}^\a (g_1+ig_2)=0~,
\cr
&&(g_1+ig_2) \check \Omega_{A,\bar\a\bar\b}-{1\over2} (g_1-i g_2) \check \Omega_{A, \g\d} \epsilon^{\g\d}{}_{\bar\a\bar\b}=0~,
\cr
&&\hat\Omega_{A,\bar\a\bar\b}-{1\over2} \hat\Omega_{A,\g\d} \epsilon^{\g\d}{}_{\bar\a\bar\b}=0~,
\cr
&&\partial_+\Phi=0~,~~~~\partial_{\bar\a}\Phi-{1\over2} H_{\bar\a\b}{}^\b+{1\over6} H_{\b_1\b_2\b_3} \epsilon^{\b_1\b_2\b_3}{}_{\bar\a}=0~,
\cr
&& g_1 \partial_{\bar\a}\Phi+{i\over2} g_2 H_{\bar\a\b}{}^\b=0~,
\la{solsyst}
\eea
where for simplicity we have chosen the gauge $f=1$ and $\a,\b,\g,\d=1,2,3,4$. This gauge can always be attained using a local $Spin(9,1)$
transformation in the direction of $\Gamma_{-+}$.

Next observe that the condition $\partial_A (g_1+ig_2)+\Omega_{A,\a}{}^\a (g_1+ig_2)=0$ implies that
$\partial_A (g_1^2+g_2^2)=0$. Since the Killing spinors are determined up to an overall constant,  we can set $g_1^2+g_2^2=1$.
In turn, we can write
\bea
i\partial_A \l+\Omega_{A,\a}{}^\a=0~,~~~      g_1+i g_2= e^{i \l}~.
\eea
Using this and (\ref{solsyst}), we find that some of the components of the fluxes can be expressed
in terms of the geometry as
\bea
&&H_{+AB}=0~,~~~H_{\b_1\b_2\b_3} \epsilon^{\b_1\b_2\b_3}{}_{\bar\a}=-6\partial_{\bar\a}(\Phi+i \l)~,~~~ H_{A\a}{}^\a=-2i\partial_A \l~,
\cr
&&\partial_{\bar\a} [\Phi-\log \cos \l]=0~,~~~\partial_+\Phi=0~,
\cr
&&H_{A'\bar\a\bar\b}={2\over e^{2i\l}-1}[ -(1+e^{2i\l})\Omega_{A', \bar\a\bar\b}+ \Omega_{A', \b_1\b_2}
\epsilon^{\b_1\b_2}{}_{\bar\a\bar\b}]~,~~~A'=-, \g~.
\la{fluxeqn}
\eea
In addition, we find the conditions
\bea
&&\Omega_{A,+B}=0~,~~~i\partial_A \l+\Omega_{A,\a}{}^\a=0~,~~~\partial_+\l=0~,~~~\Omega_{+,\a\b}=0~,
\cr
&&-(1+e^{2i\l})
[2\Omega_{\bar\b_1,\bar\b_2\bar\b_3}-\Omega_{\bar\b_3,
\bar\b_1\bar\b_2}-\Omega_{\bar\b_2, \bar\b_3\bar\b_1}] +2
\Omega_{\bar\b_1,\g\d}
\epsilon^{\g\d}{}_{\bar\b_2\bar\b_3}-\Omega_{\bar\b_3,\g\d}
\epsilon^{\g\d}{}_{\bar\b_1\bar\b_2}- \Omega_{\bar\b_2,\g\d}
\epsilon^{\g\d}{}_{\bar\b_3\bar\b_2}=0~,
\cr
&&\Omega_{\b,}{}^\b{}_{\bar\a}=-{1+2\sin^2\l\over
\sin2\l}\partial_{\bar\a}\l~,
\cr
&&\Omega_{\b_1,\b_2\b_3}\epsilon^{\b_1\b_2\b_3}{}_{\bar\a}=-{e^{i\l}\over
\sin\l}\partial_{\bar\a}\l~, \la{geomeqn}
\eea
on the geometry of spacetime. If one does not choose the gauge
$f=1$, then it is easy to see that  $f^{-1} (g_1+i g_2)=e^{i\l}$ and
that $\Omega_{A,+-}$ does not vanish but is pure gauge. Otherwise
the rest of the equations are not affected.

\subsubsection{The geometry of spacetime}

The spacetime forms $\hat\alpha$ associated with the spinor
$\hat\epsilon$ have been computed in (\ref{bieone}). After an
appropriate normalization that we suppress, the remaining ones are
\bea
&&\kappa(\check\epsilon, \check\epsilon)=
e^0-e^5~,~~~\kappa(\hat\epsilon, \check\epsilon)=\cos \l\,
(e^0-e^5)~, ~~~\xi(\hat\epsilon, \check\epsilon)=\sin \l\,
(e^0-e^5)\wedge \omega~,
\cr
&&~~~~~~~~~~~~~~\tau(\check\epsilon, \check\epsilon)=(e^0-e^5)\wedge
[{\rm Re} (e^{2i\l}\chi)-{1\over2} \omega\wedge \omega]~,~~
\cr
&&~~~~~~~~~~~~~~\tau(\hat\epsilon, \check\epsilon)=(e^0-e^5)\wedge
{\rm Re}[ e^{i\l} (\chi-{1\over2} \omega\wedge \omega)]~,
\eea
where $\chi=(e^1+i e^6)\wedge\dots\wedge (e^4+i e^9)$ and
$\omega=-(e^1\wedge e^6+e^2\wedge e^7+e^3\wedge e^8+e^4\wedge e^9)$.
The Hermitian-type of form $\omega$ gives rise to an endomorphism
$I$ of the tangent bundle of the spacetime which can be identified
with an almost complex structure transverse to  the  light-cone
directions. The conditions on the geometry can then be written in a
covariant way as
\bea
&&\nabla_A e^-=0~,~~~R_{AB}^{\cal K}=0~,~~~\nabla_X\omega_{ij}=0~,
\cr
&&\cos\l [\nabla_i \omega_{jk}]^{{\bf 20}+\bar{\bf 20}}
+{1\over2} [[(d\omega^{2,1}+d\omega^{1,2})\cdot {\rm Re}\chi^\l]_{ijk}]^{{\bf 20}+\bar{\bf 20}}=0~,
\cr
&&d\omega^{3,0}+d\omega^{0,3}=-{1\over 2\sin \l} i_{d\l} {\rm Im} \chi^\l
\cr
&&\theta_i={1\over2} d\omega_{jki} \omega^{jk}=-2{1+2\sin^2\l\over \sin2\l} \partial_i\l~,
\la{geosu4bcon}
\eea
where $e^-={1\over\sqrt{2}}(-e^0+e^5)$,  $\chi^\l=e^{i\l}\chi$,  $d\omega^{3,0}$ denotes the (3,0)-part
of $d\omega$ with respect to $I$ and similarly for $d\omega^{0,3}$, $d\omega^{2,1}$ and $d\omega^{1,2}$,
and $i,j,k=1,2,3,4,6,7,8,9$. We have also used the
decomposition of tensors in $SU(4)$ representations and
\bea
(d\omega^{2,1} \cdot {\rm Re}\chi^\l)_{ijk}={1\over2} (d\omega^{2,1})_{imn} (\chi^\l)^{mn}{}_{jk}~.
\eea
The condition $R_{AB}^{\cal K}=0$ implies that the curvature of the
canonical bundle of the spacetime vanishes and this is derived as an
integrability condition of $i\partial_A\l+\Omega_{A,\a}{}^\a=0$. The
one-form $e^-$ is parallel with respect to the Levi-Civita
connection. Therefore the associated vector field $X=e_+$ is null,
Killing and rotation free. Adapting coordinates along $X$,
$X=\partial/\partial u$, and using some of the conditions, the
metric and three-form $H$ can be written as
\bea
&&ds^2= 2 e^- e^++ \delta_{ij} e^i e^j= 2 dv (du+ Vdv+n)+ g_{IJ} dy^I dy^J~,
\cr
&&H={1\over2} i_{d\Phi-d_I\lambda} {\rm Re}\chi-\cot\l [(d\omega)^{1,2}+ (d\omega)^{2,1}]-{1\over2 \sin\l}
 ({1\over3}W_1+ W_2)\cdot{\rm Re}\chi^\l
\cr
&&~~~~-e^-\wedge [\cot\l\,\nabla_-\omega+{1\over2 \sin \l}\,\nabla_-\omega\cdot {\rm Re}\chi^\l]
+{1\over2} e^-\wedge\omega\, \partial_-\l +{1\over2} H^{\bf 15}_{-ij} e^-\wedge e^i\wedge e^j~,
\cr
&&e^{\Phi}= \ell(v) \cos \l~,
\eea
where
\bea
(W_1\cdot {\rm Re}\chi^\l)_{\a,\bar\g\bar\d}={1\over2}(W_1)_{\a\b_1\b_2} {\rm Re}(\chi^\l)^{\b_1\b_2}{}_{\bar\g\bar\d}~,
\eea
the classes $W_1$ and $W_2$ are defined in appendix C and $d_I$ is
the exterior derivative with respect to the endomorphism $I$.
 The only component of $H$ that
is not determined in terms of the geometry is
$H^{\mathfrak{su}(4)}_{-ij}=H^{\bf 15}_{-ij}$. The expression for
the flux $H$ depends on the trivialization of the canonical bundle
$\l$. All the components of the metric and fluxes are independent of
the coordinate $u$.

The spacetime is a pp-wave propagating in a manifold $B$ with an
$SU(4)$ structure given by $v,u={\rm const}$. Alternatively, it can
be seen as a Lorentzian two-parameter family of $B$. The geometry of
$B$ can be easily described by restricting the conditions we have
presented in (\ref{geosu4bcon}) on $B$. In particular we have that
$\nabla_A \chi^\l=0$
 implies that
\bea
2\tilde W_5^\lambda-\tilde W_4^\lambda=0~,
\la{bconsua1}
\eea
where $\tilde W_5$ denotes the restriction of the forms that define the class $W_5$ on $B$ and similarly for the rest.
In addition, $\tilde W_1$ and $\tilde W_4$ are specified in terms of $d\lambda$  as
\bea
\tilde W_1=-{1\over2 \sin \l}  i_{ d\lambda} {\rm Im}\chi^\l~,
\cr
\tilde W_4=-2{1+2\sin^2\l\over \sin2\l} d\l~.
\la{bconsua2}
\eea
Finally, we have
\bea
2\cos\l\, \tilde W_2+{1\over2} [\tilde W_3\cdot{\rm Re} \chi^\l]^{{\bf 20}+\bar{\bf 20}}=0~.
\la{bconsua3}
\eea
Therefore the eight-manifold $B$  is almost hermitian with trivial
canonical bundle.

\subsection{Backgrounds with $G_2$-invariant Killing spinors}

\subsubsection{Supersymmetry conditions}

We have seen that the $G_2$-invariant Killing spinors can be chosen
as
\bea
\hat\epsilon=f(1+e_{1234})~,~~~\check\epsilon=g (e_{15}+e_{2345})~.
\eea
The local $Spin(9,1)$ transformations along the $\Gamma_{+-}$
direction scale  the Killing spinors as $\hat\epsilon\rightarrow
\ell \hat\epsilon$ and $\check\epsilon\rightarrow \ell^{-1}
\check\epsilon$. Therefore one can  choose the gauge $f=g$. In what
follows, we shall present our results in this gauge.  Of course the
geometry of spacetime is independent of the choice of gauge.

The linear system  for the above $G_2$-invariant spinors
can be easily constructed from that
of the heterotic backgrounds with $G_2$-invariant Killing spinors \cite{ugplgp}. The only difference is that
one has to replace $H$ with $-H$ in the conditions that arise from the second Killing spinor $\check\epsilon$.
Because of this, we shall not give further  details. After some computation, the conditions
that arise from the gravitino Killing spinor equations
can be written as
\bea
2\partial_A\log f^2- H_{A-+}=0~,~~~\hat\nabla_A(f^2 e^-)=0~,~~~\check\nabla_A(f^2 e^+)=0~,
\cr
H_{A1i}+{1\over12} \nabla_A\varphi_{jkl} \star\varphi^{jkl}{}_i =0~,~~~\nabla_AZ_i-{1\over4} H_{Ajk} \varphi^{jk}{}_i=0~,
\la{gravg2b}
\eea
and the conditions that arise from the dilatino Killing spinor equation are
\bea
&&\partial_+\Phi=\partial_-\Phi=0~,~~~H_{-1i}+{1\over2} H_{-kl} \varphi^{kl}{}_i=0~,~~~H_{+1i}-{1\over2} H_{+kl} \varphi^{kl}{}_i=0~,
\cr
&&H_{jkl}\star\varphi^{jkl}{}_i=0~,~~~2 \partial _1\Phi-{1\over6} H_{ijk} \varphi^{ijk}- H_{-+1}=0~,
\cr
&&2\partial_i\Phi+{1\over2} H_{1kl} \varphi^{kl}{}_i-H_{-+i}=0~,
\la{dilg2b}
\eea
where $i,j,k=2,3,4,6,7,8,9$, $\varphi={\rm Re}\, \hat\chi+ e^6\wedge
\hat\omega$ is the $G_2$-invariant three-form, $\hat\chi= (e^2+i
e^7)\wedge (e^3+i e^8)\wedge (e^4+i e^9)$, $\hat\omega=-(e^2\wedge
e^7+e^3\wedge e^8+e^4\wedge e^9)$ and $Z=e_1$. The dual
$\star\varphi$ of $\varphi$ is taken with respect to $d{\rm
vol}=e^2\wedge e^3\wedge e^4\wedge e^6\wedge e^7\wedge e^8\wedge
e^9$.

\subsubsection{The geometry of spacetime}

The spacetime form bilinears of the spinor $\hat\epsilon$ have been presented in (\ref{bieone}). The rest of the
form bilinears are
\bea
&&\kappa(\hat\epsilon, \check\epsilon)=- f^2 e^1~,~~~\kappa(\check\epsilon, \check\epsilon)= f^2 (e^0+ e^5)~,~~~
\xi(\hat\epsilon, \check\epsilon)=f^2[{\rm Re}\hat\chi+ e^6\wedge
\hat\omega-e^0\wedge e^1\wedge e^5]~,
\cr
&&\tau(\hat\epsilon, \check\epsilon)= f^2 [-{\rm Re} \hat\chi\wedge e^0\wedge e^5+{\rm Im}\hat\chi\wedge e^1\wedge e^6+{1\over2} e^1\wedge \hat\omega\wedge \hat\omega-
\hat\omega\wedge e^0\wedge e^5\wedge e^6]~,
\cr
&&\tau(\check\epsilon, \check\epsilon)=- f^2 (e^0+ e^5)\wedge [  e^1\wedge {\rm Re}\hat\chi+ e^6\wedge {\rm Im} \hat\chi+{1\over2} \hat\omega\wedge \hat\omega
+\hat\omega\wedge e^1\wedge e^6]~.
\eea
{}It is clear from (\ref{gravg2b})  that $\kappa(\hat\epsilon,
\hat\epsilon)$ and $\kappa(\check\epsilon, \check\epsilon)$ are
parallel with respect to the $\hat\nabla$ and $\check\nabla$
connections, respectively, as may have been expected from the
general arguments we have presented in section two. In fact, a more
detailed analysis reveals that ${\rm hol}(\hat\nabla)\subseteq
Spin(7)\ltimes\bR^8$ and ${\rm hol}(\check\nabla)\subseteq
Spin(7)\ltimes\bR^8$. These  holonomy groups are embedded  in
$Spin(9,1)$ in different ways. This can also be seen by comparing
$\tau (\hat\epsilon, \hat\epsilon)$ and $\tau(\check\epsilon,
\check\epsilon)$.

To solve the conditions (\ref{gravg2b}) and (\ref{dilg2b}) observe that
 under $G_2$  the flux $H$ decomposes as
\bea
H_{abc}~,~~~H_{abi}~,~~~H_{aij}~,~~~H_{ijk}~,~~~a,b,c=+,-,1~,~~~i,j,k=2,3,4,6,7,8,9~.
\eea
In addition the space of two- and three-forms decomposes as
$\Lambda^2(\bR^7)=\Lambda^2_{\bf 7}\oplus \Lambda^2_{\bf 14}$ and
$\Lambda^3(\bR^7)=\Lambda^3_{\bf 1}\oplus \Lambda^3_{\bf 7}\oplus
\Lambda^3_{\bf 27}$, respectively, where $\Lambda^2_{\bf
14}=\mathfrak{g}_2$. The conditions (\ref{gravg2b}) and
(\ref{dilg2b})  determine all the components of $H$ in terms of the
geometry
 apart from $H_{-ij}^{\bf 14}$
and $H_{+ij}^{\bf 14}$. Let $\hat X, \check X$ and $Z$ be the vector
fields associated with the bi-linears $f^2 e^-$, $f^2 e^+$ and
$e^1$, i.e.~$\hat X=f^2 e_+$, $\check X=f^2 e_-$ and $Z=e_1$. One
then finds that
\bea
&&2\partial_1\log f^2-H_{1-+}=0~,~~H_{-1i}+2\nabla_-
Z_i=0~,~~~H_{+1i}-2\nabla_+ Z_i=0~,
\cr
&&2\partial_i\log f^2-H_{i-+}=0~,~~~H_{1ij}+{1\over12}
\nabla_{[i}\varphi^{mnp} \star \varphi_{j]mnp}=0~,~~~
\nabla_+Z_i-{1\over4} H_{+jk}\varphi^{jk}{}_i=0~,
\cr
&&\nabla_-Z_i-{1\over4}
H_{-jk}\varphi^{jk}{}_i=0~,~~H_{ijk}=-{1\over3} \nabla_lZ^l
\varphi_{ijk}+ 3 Z_{p[i}
\varphi^p{}_{jk]}~,~~\partial_-\Phi=\partial_+\Phi=0~, \la{fg2b}
\eea
where $Z_{ij}=\nabla_{(i} Z_{j)}$.
In addition, the conditions on the geometry are
\bea
&&\hat\nabla_AX^B=\check\nabla_AY^B=0~,~~~\nabla_1\varphi_{jkl} \star\varphi^{jkl}{}_i=0~,~~~\nabla_{(i}\varphi^{mnp}
\star\varphi_{j)mnp}=0~,
\cr
&&8\nabla_1Z_i+\nabla^p\varphi_{pmn} \varphi^{mn}{}_i=0~,~~~{1\over24}\nabla_-\varphi_{jkl}\star\varphi^{jkl}{}_i+\nabla_-Z_i=0~,
\cr
&&{1\over24}\nabla_+\varphi_{jkl}\star\varphi^{jkl}{}_i-\nabla_+Z_i=0~,~~~\nabla_{[i} Z_{j]}=0~,
\cr
&&\partial_1(\Phi-\log f^2)-{1\over3} \nabla_i Z^i=0~,~~~2\partial_i(\Phi-\log f^2)-{1\over4} \nabla^p\varphi_{pjk} \varphi^{jk}{}_i=0~.
\la{gg2b}
\eea

As we have already mentioned, the conditions on the geometry imply
that $\hat X,\check X$ are parallel with respect to the $\hat\nabla$
and $\check\nabla$ connections, respectively. This implies that
$\hat X,\check X$ are Killing, commute $[\hat X,\check X]=0$ and
their rotations are given in terms of $H$ as $d\hat\kappa_X=i_{\hat
X}H$ and $d\check\kappa_Y=-i_{\check Y}H$, see section two. In
addition, it turns out that $[\hat X,Z]=[\check X,Z]=0$ but $Z$ is
not Killing. Because of this, it is convenient to consider the
spacetime as a fibration over an eight-dimensional space $B$ with
fibres given by the orbits of $\hat X,\check X$. The base space $B$
admits a $G_2$-structure which we shall specify.
 Introducing coordinates adapted to
the vectors fields $\hat X,\check X$ and $Z$, i.e.~$\hat
X=\partial/\partial u$, $\check X=\partial/\partial v$ and
$Z=\partial/\partial x$, one can write the metric and the three-form
as
\bea
&&ds^2=  2 f^4 (du+ n_i e^i) (dv+m_i e^i)+ (dx+\ell_i e^i)^2+ \delta_{ij} e^i{}_I e^j{}_J
dy^I dy^J
\cr
&&H=d(e^-\wedge e^+)-{1\over24} \nabla_{[i}\varphi^{mnp} \star \varphi_{j]mnp}\, \,e^1\wedge e^i\wedge e^j
\cr
&&~~~~~~+[-{1\over18} \nabla_lZ^l \varphi_{ijk}+ {1\over2} Z_{p[i} \varphi^p{}_{jk]}]
\, e^i\wedge e^j\wedge e^k~,
\cr
&&\Phi=\Phi(x, y)
\eea
where
\bea
e^+= f^2(du+ n_i e^i)~,~~~~e^-= f^2 (dv+m_i e^i)~,~~~e^1=dx+\ell_i e^i~,
\eea
and all components of the metric and fluxes are independent of
$u,v$. The components $m,n$ of the metric are not arbitrary. In
particular, they are related to the rotation of $\hat X,\check X$
which in turn is related to $Z$ as can be seen by the conditions in
(\ref{fg2b}).

It remains to find the conditions that supersymmetry imposes on the eight-dimensional base space  $B$
of the fibration. $B$ admits a $G_2$-structure. The different $G_2$-structures of an eight-dimensional
manifold are described in appendix C.
It is clear that the conditions (\ref{gg2b}) imply that
\bea
&&\tilde W=0,~~~ \tilde X_2=\tilde X_3=0~,~~~\tilde
W_4=0~,~~~4\tilde X+3\tilde W_2=0~,
\cr
&&3\partial_1(\Phi-\log f^2)-\tilde X_1=0~,~~~4\partial_i(\Phi-\log
f^2)-3\tilde W_2=0~, \la{bcong2b}
\eea
where $\tilde W_2$ is represented by the Lee form\footnote{In terms of seven-dimensional data $\tilde\theta=-{1\over3} \star
(\star d\varphi\wedge\varphi)$.}
\bea
\tilde \theta_i={1\over6} \nabla^p\varphi_{pmn} \varphi^{mn}{}_i~,
\eea
 $\tilde W$ denotes the projection on the base space $B$ of $W$ and similarly for the rest of the classes.
The components $H_{ijk}$ are determined by the classes $\tilde X_1$ and $\tilde X_4$ while the component $H_{1ij}$ is determined
by $\tilde W_2$ and $\tilde W_3$. Observe that if $e^{\Phi}=f^2$, then $\tilde W_2=\tilde X=\tilde X_1=0$.

\subsubsection{Field equations}

To find the field equations that are implied by the Killing spinor equations, one can use
the integrability conditions in appendix A and the results in appendix B. A straightforward calculation
 reveals that if  one imposes
\bea
dH=0~,~~~LH_{-+}=0~,
\eea
then all the rest of the field equations are implied. This is the
case for all the $\hat N=\check N=1$ supersymmetric common sector
backgrounds which admit $K$-invariant Killing spinors. As the proof
is very similar for the other cases we shall not repeat the analysis
in each case.

\newsection{IIA $N=2$ backgrounds}

\subsection{Backgrounds with $Spin(7)$-invariant spinors}

We have shown in section two that the $Spin(7)$-invariant Killing spinors can be written as
\bea
\hat\epsilon= f(1+e_{1234})~,~~~\check\epsilon=g (e_5+e_{12345})~.
\eea
Observe that $\hat\epsilon$ is an even-degree form while
$\check\epsilon$ is an odd-degree form. This is because the Killing
spinors of the two copies of the IIA Killing spinor equations have
opposite chirality.  The linear system associated with these Killing
spinors can be constructed from that of \cite{ugplgp} and the
results in appendix B. Under local $Spin(9,1)$ transformations in
the direction $\Gamma_{+-}$, the spinors transform as
$\hat\epsilon\rightarrow \ell\hat\epsilon$ and
$\check\epsilon\rightarrow \ell^{-1}\check\epsilon$. This symmetry
can be fixed by setting $f=g$. It turns out that this is a
convenient gauge to use for our investigation. The geometry of
spacetime does not depend on the choice of gauge.

The linear system associated with the IIA Killing spinor equations for the $\hat\epsilon$ and $\check \epsilon$
 spinors can be solved to give
\bea
&&\Omega_{A,-+}=0~,~~~2\partial_A\log (f^2)+H_{A+-}=0~,~~~\Omega_{A,\a}{}^\a=0~,
\cr
&&\Omega_{A,\bar\a\bar\b}-{1\over2} \Omega_{A, \g\d} \epsilon^{\g\d}{}_{\bar\a\bar \b}=0~,~~~\hat\Omega_{A,+\a}=0~,
~~~\check\Omega_{A,-\a}=0~,
\cr
&&\partial_+\Phi=0~,~~~H_{+\a}{}^\a=0~,~~~-H_{+\bar\a\bar\b}+{1\over2} H_{+\g\d} \epsilon^{\g\d}{}_{\bar\a\bar \b}=0~,
\cr
&&\partial_-\Phi=0~,~~~H_{-\a}{}^\a=0~,~~~-H_{-\bar\a\bar\b}+{1\over2} H_{-\g\d} \epsilon^{\g\d}{}_{\bar\a\bar \b}=0~,
\cr
&&H_{ijk}=0~,~~~\partial_{\bar\a}\Phi-{1\over2} H_{-+\bar\a}=0~.
\eea
These conditions can be rewritten in a covariant way as
\bea
&&\hat\nabla (f^2 e^-)=0~,~~~\check\nabla (f^2 e^+)=0~,~~~\nabla_A\phi_{ijkl}=0~,~~~
\cr
&&\partial_+\Phi=\partial_-\Phi=0~,~~~\partial_+f^2=\partial_-f^2=0~,~~~
\cr
&&-H_{-ij}+{1\over 2} H_{-kl} \phi^{kl}{}_{ij}=0~,~~~-H_{+ij}+{1\over 2} H_{+kl} \phi^{kl}{}_{ij}=0~,~~~
\cr
&&H_{ijk}=0~,~~~\partial_i\Phi-{1\over2} H_{-+i}=0~,
\la{conspin7a}
\eea
where $\phi$ is the $Spin(7)$-invariant four-form defined in section three.
It remains to examine the restrictions on the geometry of spacetime imposed by the above conditions.

\subsubsection{The geometry of spacetime}

The spinor bilinears of $\hat\epsilon$ have already been computed in
the IIB case. The spinor bi-linears of $\check\epsilon$ can be
computed from those of $\hat\epsilon$ by replacing $e^-$ with $e^+$.
It remains to compute the spacetime forms associated with
$(\hat\epsilon, \check\epsilon)$. After an additional normalization
of the spinors with $1/\sqrt 2$, one finds
\bea
\alpha(\hat\epsilon, \check\epsilon)=-f^2~,~~~\beta(\hat\epsilon, \check\epsilon)=f^2\, e^0\wedge e^5~,~~~
\rho(\hat\epsilon, \check\epsilon)= -f^2\,\phi~,
\eea
where $\phi$ is the $Spin(7)$-invariant form in section three. In
contrast to the IIB case, the Killing spinors have a non-degenerate
inner product.

A consequence of the supersymmetry conditions (\ref{conspin7a}) is
that the vector fields $\hat X=f^2 e_+$ and $\check X= f^2 e_-$ are
Killing and commute $[\hat X,\check X]=0$. Adapting coordinates
along $\hat X$ and $\check X$, $\hat X=\partial/\partial u$, $\check
X=\partial/\partial v$, the metric, torsion and dilaton can be
written as
\bea
&&ds^2=2 f^4 (dv+m) (du+n)+ \delta_{ij} e^i e^j~,
\cr
&&H=d(e^-\wedge e^+)~,
\cr
&&e^\Phi= f^2~,
\eea
where $e^-=f^2 (dv+m)$ and $e^+= f^2 (du+n)$, and  all fields are
independent of $u,v$. The components $m, n$ are not arbitrary. In
particular, the conditions on $H$ in (\ref{conspin7a}) require that
  $dm$ and $dn$ take values in  $\mathfrak{spin}(7)$.
Clearly the spacetime is a rank two fibre bundle with fibre given by
the orbits of $\hat X$ and $\check X$
 and with base  space $B$  a $Spin(7)$ manifold,
 i.e.~${\rm hol} (\tilde\nabla)\subseteq Spin(7)$.

 The Bianchi identity for $H$ is automatically satisfied. So, as we have explained in section three,
  it remains to impose the field equation
 $LH_{-+}=0$. This specifies  $f$. In particular, it implies
 that $f^{-4}$ is a harmonic function on $B$. Therefore the spacetime can be thought of as a generalization of
 a fundamental string of \cite{ruiz} with rotation and wrapping, and with transverse space the  $Spin(7)$ manifold $B$.

\subsection{Backgrounds with $SU(4)$-invariant spinors}

It has been shown in section two that the Killing spinors can be chosen as
\bea
\hat\epsilon= f(1+e_{1234})~,~~~\check\epsilon=g_1 (e_5+e_{12345})+i g_2 (e_5-e_{12345})~.
\eea
The linear system associated with the Killing spinor equations for these spinors can be easily
constructed from the results of \cite{ugplgp} and those in appendix B. So we shall not give more details.

It is convenient to express the solution of the linear system in the
gauge $f=1$. This is attained with a local $Spin(9,1)$
transformation in the $\Gamma_{+-}$ direction. After some
computation, and setting  $g_1+g_2= g^2 e^{i\l}$, we find that the
solution of the linear system can be written as
\bea
&&\Omega_{A,\a}{}^\a={1\over2}H_{A\a}{}^\a=-i\partial_A\l~,~~~\hat\Omega_{A,+B}=0~,~~~\partial_A\log g^2-\Omega_{A,-+}=0~,
~~~
\check\Omega_{A,-\a}=0~,
\cr
&&\partial_+\Phi=\partial_-\Phi=0~,~~~H_{-\a}{}^\a=H_{+\a}{}^\a=0~,~~~e^\Phi=g^2\cos\l~,~~~
\cr
&&H_{\b_1\b_2\b_3} \epsilon^{\b_1\b_2\b_3}{}_{\bar\a}=-6 \partial_{\bar\a}(\log\cos\l+i \l)~,
\cr
&&H_{A'\bar\a\bar\b}={2\over e^{2i\l}-1}[ -(1+e^{2i\l})\Omega_{A', \bar\a\bar\b}+ \Omega_{A', \b_1\b_2}
\epsilon^{\b_1\b_2}{}_{\bar\a\bar\b}]~,~~~A'=\bar\g, \g~.
\cr
&&-(1+e^{2i\l})(2\Omega_{\bar\a,\bar\b\bar\g}-\Omega_{\bar\g,\bar\a\bar\b}-\Omega_{\bar\b,\bar\g\bar\a})
+2\Omega_{\bar\a,\d_1\d_2} \epsilon^{\d_1\d_2}{}_{\bar\b\bar\g}-\Omega_{\bar\g,\d_1\d_2} \epsilon^{\d_1\d_2}{}_{\bar\a\bar\b}-
\Omega_{\bar\b,\d_1\d_2} \epsilon^{\d_1\d_2}{}_{\bar\g\bar\a}=0~,
\cr
&&-2\Omega_{+,\bar\a\bar\b}+\Omega_{+,\g\d}\epsilon^{\g\d}{}_{\bar\a\bar\b}=0~, -2\Omega_{-,\bar\a\bar\b}+ e^{-2i\l}
\Omega_{-,\g\d}\epsilon^{\g\d}{}_{\bar\a\bar\b}=0~,~~~
\cr
&&\Omega_{-,\a\b}-{1\over2} H_{-\a\b}=0~,~~~\Omega_{+,\a\b}+{1\over2}H_{+\a\b}=0~.
\eea
In turn, these conditions can be solved to express  the fluxes in
terms of the geometry
\bea
{1\over2}H_{A\a}{}^\a=-i\partial_A\l~,~~~H_{-\a}{}^\a=H_{+\a}{}^\a=0~,~~~e^\Phi=g^2 \cos\l~,
\cr
H_{\b_1\b_2\b_3} \epsilon^{\b_1\b_2\b_3}{}_{\bar\a}=-6 \partial_{\bar\a}(\log\cos\l+i \l)~,
\cr
H_{\g\bar\a\bar\b}={2\over e^{2i\l}-1}[ -(1+e^{2i\l})\Omega_{\g, \bar\a\bar\b}+ \Omega_{\g, \b_1\b_2}
\epsilon^{\b_1\b_2}{}_{\bar\a\bar\b}]~,
\cr
\Omega_{-,\a\b}-{1\over2} H_{-\a\b}=0~,~~~\Omega_{+\a\b}+{1\over2}H_{+\a\b}=0~,
\eea
and to find  the conditions
\bea
&&\hat\Omega_{A,+B}=0~,~~~i\partial_A\l+\Omega_{A,\a}{}^\a=0~,~~~\partial_A\log g^2-\Omega_{A,-+}=0~,
\cr
&&\check\Omega_{A,-\a}=0~,~~~-2\Omega_{+,\bar\a\bar\b}+\Omega_{+,\g\d}\epsilon^{\g\d}{}_{\bar\a\bar\b}=0~,
-2\Omega_{-,\bar\a\bar\b}+ e^{-2i\l}
\Omega_{-,\g\d}\epsilon^{\g\d}{}_{\bar\a\bar\b}=0~,~~~
\cr
&&-(1+e^{2i\l})(2\Omega_{\bar\a,\bar\b\bar\g}-\Omega_{\bar\g,\bar\a\bar\b}-\Omega_{\bar\b,\bar\g\bar\a})
+2\Omega_{\bar\a,\d_1\d_2} \epsilon^{\d_1\d_2}{}_{\bar\b\bar\g}-\Omega_{\bar\g,\d_1\d_2} \epsilon^{\d_1\d_2}{}_{\bar\a\bar\b}-
\Omega_{\bar\b,\d_1\d_2} \epsilon^{\d_1\d_2}{}_{\bar\g\bar\a}=0~,
\cr
&&\Omega_{\b,}{}^\b{}_{\bar\a}=-{1+2\sin^2\l\over \sin2\l}\partial_{\bar\a}\l~,
\cr
&&\Omega_{\b_1,\b_2\b_3}\epsilon^{\b_1\b_2\b_3}{}_{\bar\a}=-{e^{i\l}\over \sin\l}\partial_{\bar\a}\l~,
\eea
on the geometry of spacetime. It is clear that the conditions we
have found resemble those of the $SU(4)\ltimes\bR^8$ backgrounds of
the type IIB common sector.

\subsubsection{The geometry of spacetime}

The spacetime forms $\hat\a$ associated with the spinor
$\hat\epsilon$ have been computed in (\ref{bieone}). Similarly, the
spacetime forms of $\check\epsilon$ can be easily constructed from
those of the second spinor in the $SU(4)\ltimes\bR^8$ case.  The
remaining spacetime form spinor bilinears are
\bea
\alpha(\hat\epsilon, \check\epsilon)=- g_1~,~~~\beta(\hat\epsilon,
\check\epsilon)=  g_1 e^0\wedge e^5- g_2 \omega~,~~~~~~~\cr
\rho(\hat\epsilon, \check\epsilon)= g_2 e^0\wedge e^5\wedge\omega+
\frac{1}{2} g_1 \omega\wedge\omega -  g_1 {\rm Re} \chi +  g_2
{\rm Im} \chi~,
\eea
where $\omega$ and $\chi$ are defined as in section three.
The conditions on the geometry can now be written in a covariant way as
\bea
&&\hat\nabla_A e^-=0~,~~~\check\nabla_A (g^2 e^+)=0~,~~~R_{AB}^{\cal K}=0~,~~~\check\nabla_+\omega_{ij}=0~,~~~\hat\nabla_-\omega_{ij}=0~,
\cr
&&\nabla_+\phi=0~,~~~\nabla_-\phi^{2\l}=0~,
\cr
&&\cos\l [\nabla_i \omega_{jk}]^{{\bf 20}+\bar{\bf 20}}
+{1\over2} [[(d\omega^{2,1}+d\omega^{1,2})\cdot {\rm Re}\chi^\l]_{ijk}]^{{\bf 20}+\bar{\bf 20}}=0~,
\cr
&&d\omega^{3,0}+d\omega^{0,3}=-{1\over 2\sin \l} i_{d\l} {\rm Im} \chi^\l
\cr
&&\theta_i={1\over2} d\omega_{jki} \omega^{jk}=-2{1+2\sin^2\l\over \sin2\l} \partial_i\l~,
\la{geosu4bconx}
\eea
where  $R^{\cal K}$ is the curvature of the canonical bundle whose
condition arises as the integrability condition of
$i\partial_A\l+\Omega_{A,\a}{}^\a=0$ and $\phi^{2\l}={\rm Re}
(e^{2i\l} \chi)-{1\over 2} \omega\wedge\omega$. The explanation for
the rest of the notation can be found in  the type IIB
$SU(4)\ltimes\bR^8$ case in section three.

The conditions on the geometry imply that the vector fields $\hat
X=e_+$ and $\check X=g^2 e_-$ are $\hat\nabla$- and $\check\nabla$-
parallel, respectively. Therefore they are Killing and $[\hat X,
\check X]=0$. After adapting coordinates along these Killing vector
fields, $\hat X=\partial/\partial u$ and $\check X=\partial/\partial
v$, and after some computation, the spacetime metric  torsion and
dilaton can be written as
\bea
&&ds^2=2f^4 (dv+n)(du+m)+ \delta_{ij} e^i_I e^j_J dy^I dy^J
\cr
&& H =d (e^-\wedge e^+)+{1\over2} i_{d\Phi-d_I\lambda} {\rm Re}\chi-\cot\l [(d\omega)^{1,2}+ (d\omega)^{2,1}]-{1\over2 \sin\l}
 ({1\over3}W_1+ W_2)\cdot{\rm Re}\chi^\l
 \cr
 &&
\Phi=f^4(y) \cos\l(y)~,
\eea
where we have set $g=f^2$ and all the fields are independent of the
coordinates $u,v$.

The spacetime is a fibre bundle with fibre given by the orbits of
$\hat X$ and $\check X$ and with base space
 an eight-dimensional space $B$. The geometry of $B$ can be easily described using (\ref{geosu4bconx}). It turns out
 that one finds the {\it same} conditions as those on the transverse space $B$ of the pp-wave spacetime of
 $N=2$ IIB backgrounds with    $SU(4)\ltimes \bR^8$-invariant spinors  (\ref{bconsua1})-(\ref{bconsua3}).

\subsection{Backgrounds with $G_2\ltimes \bR^8$-invariant spinors}

\subsubsection{Supersymmetry conditions}

As we have demonstrated in section two, the  Killing spinors can be chosen as,
\bea
\hat\epsilon&=&f (1+e_{1234})~,~~~
\check\epsilon= g (e_{1}+ e_{234})~.
\la{kspingtwo}
\eea
The linear system associated with the Killing spinor equations for these spinors can be
easily constructed using the results of \cite{ugplgp} and those in appendix B.
To solve the linear system, it is convenient to work in the gauge $f=1$ which as we have explained is attained
by a local $Spin(9,1)$ transformation. Then
after some computation, the solution of the linear system can be written as
\bea
&&\nabla_A e^-=0~,~~~{1\over12} \nabla_A\varphi_{jkl} \star\varphi^{jkl}{}_i+ H_{A1i}=0~,
\cr
&&\nabla_A Z_i-{1\over4} H_{Ajk} \varphi^{jk}{}_i=0~,~~~g=f=1~,
\cr
&&\partial_+\Phi=0~,~~~ H_{jkl}\star\varphi^{jkl}{}_i=0~, H_{+AB}=0~,
\cr
&&2\partial_1\Phi-{1\over6} H_{jkl}\varphi^{jkl}=0~,~~~2\partial_i\Phi+{1\over2} H_{1jk} \varphi^{jk}{}_i=0~,
\la{cong2a}
\eea
where $Z=e_1$.  The rest of the notation is described in the $G_2$
case of section three where one can also find the definition of the
$G_2$-invariant three-form $\varphi$. It remains to solve these
conditions to determine the restrictions on the geometry and fluxes.

\subsubsection{The geometry of spacetime}

The form spinor bilinears of $(\hat\epsilon, \hat\epsilon)$ have already been computed. The form bilinears
of $(\check\epsilon, \check\epsilon)$ can be computed from those of $(e_{51}+e_{5234}, e_{51}+e_{5234})$
by replacing $e^-$ with $e^+$. So the only new form spinor bilinears are those of $(\hat\epsilon, \check\epsilon)$.
It is easy to see after an additional normalization of the spinors that
\bea
\alpha(\hat\epsilon, \check\epsilon)=0~,~~~\beta(\hat\epsilon, \check\epsilon)= (e^0-e^5)\wedge e^1~,~~~
\rho(\hat\epsilon, \check\epsilon)= -(e^0-e^5)\wedge \varphi~.
\eea
The supersymmetry conditions (\ref{cong2a}) can be solved to express
some of the fluxes in terms of the geometry
\bea
&&H_{+AB}=0~,~~H_{-1i}=-{1\over12}\nabla_-\varphi_{jkl} \star\varphi^{jkl}{}_i~,~~
H^{\bf 7}_{-ij}={2\over3}\nabla_-Z_k \varphi^k{}_{ij}~,~~~\partial_+\Phi=0~,
\cr
&&H_{1ij}=-{1\over12} \nabla_{[i}\varphi^{klm} \star\varphi_{j]klm}~,~~~H_{ijk}=-{1\over3} \nabla_l Z^l \varphi_{ijk}
+3 Z_{p[i} \varphi^p{}_{jk]}~,
\eea
and to find the  conditions
\bea
&&\nabla_A e^-=0~, \nabla_1\varphi_{jkl} \star \varphi^{jkl}{}_i=0~,~~~\nabla_+\varphi_{jkl} \star \varphi^{jkl}{}_i=0~,~~
\cr
&&\nabla_{(i}\varphi^{jkl} \star \varphi_{m)jkl}=0~,~~~
8 \nabla_1 Z_i+\nabla^m \varphi_{mjk} \varphi^{jk}{}_i=0~,~~~\nabla_+Z_i=0~,~~~\nabla_{[i} Z_{j]}=0~,~~~
\cr
&&\partial_1\Phi-{1\over3} \nabla_i Z^i=0~,~~~2\partial_i\Phi-{1\over4} \nabla^l \varphi_{ljk} \varphi^{jk}{}_i=0~,
\eea
on the geometry of spacetime.
It is clear from the above geometric conditions that the null vector field
 $\hat X=e_+$ is parallel with respect to the Levi-Civita connection. This  implies that $\hat X$
 is Killing and rotation free.  Adapting coordinates along $\hat X=\partial/\partial u$,
the metric and torsion  can be written as
\bea
&&ds^2=2 dv (du+ V dv+ n)+ \delta_{ij} e^i_I e^j_J dy^I dy^J
\cr
&&H=-{1\over12}\nabla_-\varphi_{jkl} \star\varphi^{jkl}{}_i\, e^-\wedge e^1 \wedge e^i+
{1\over3}\nabla_-Z_k \varphi^k{}_{ij}\, e^-\wedge e^i\wedge e^j
\cr
&&~~~~+{1\over2} H^{\bf 14}_{-ij}\, e^-\wedge e^i\wedge e^j-{1\over24}\nabla_{[i}\varphi^{klm} \star\varphi_{j]klm}
\, e^1\wedge e^i\wedge e^j
\cr
&&~~~~+[-{1\over18} \nabla_l Z^l \varphi_{ijk}+{1\over2} Z_{p[i} \varphi^p{}_{jk]}]\, e^i\wedge e^j\wedge e^k~,
\cr
&&\Phi=\Phi(v,y)~,
\eea
where $e^-=dv$ and $e^+=(du+ V dv+ n)$, and all the fields are
independent of the coordinate $u$. Therefore the only component of
$H$ that is not determined in terms of the geometry is $H_{-ij}^{\bf
14}$.

The spacetime is a pp-wave propagating in the transverse space $B$ defined by $u,v={\rm const}$. Alternatively,
it can be interpreted as a Lorentzian deformation family of $B$. The manifold $B$ is an eight-dimensional
manifold equipped with a $G_2$-structure. The geometry of such manifolds has been described in appendix C.
It is straightforward to find the conditions that supersymmetry imposes on the $G_2$-structure of $B$.
In particular, one finds that
\bea
&&\tilde W=0~,~~~\tilde X_2=\tilde X_3=0~,~~~\tilde W_4=0~,~~~4\tilde X+3\tilde W_2=0~,~~~
\cr
&&3\partial_1\tilde \Phi-\tilde X_1=0~,~~~4\partial_i\tilde \Phi-3\tilde W_2=0~,
\eea
where $\tilde W$ denotes the restriction of $W$ on $B$ and similarly for the rest of the classes.
So up to a redefinition of the dilaton,  $B$ has the {\it same} geometry as the base space of the fibration
that arises in the $N=2$ supersymmetric backgrounds with $G_2$-invariant spinors in the type IIB common sector
(\ref{bcong2b}).

\newsection{Type II strings in $N=2$ backgrounds}

\subsection{Light-cone gauge fixing}

As an application of our results, we shall show that one can always
gauge fix the world-volume action of a string propagating in $\hat
N, \check N\geq 1$ supersymmetric backgrounds in the light-cone
gauge. Since the $\hat N, \check N\geq 1$ supersymmetric backgrounds
are special cases of the $\hat N=\check N=1$  ones, it suffices to
demonstrate this for the latter. The details of the gauge fixing
procedure  depend on whether the stability subgroup of the Killing
spinors is $K$ or $K\ltimes\bR^8$. First let us consider the latter
case. After putting the world-sheet metric in conformal gauge, the
bosonic part of the Lagrangian of a string propagating in background
(\ref{nullback}) is
\bea
L&=&\partial_{\pp}v (\partial_= u+V \partial_=v+(n_I+b_I) \partial_= y^I)+\partial_=v (\partial_{\pp} u+V \partial_{\pp}v+(n_I-b_I) \partial_{\pp} y^I)
\cr
&&+
(g_{IJ}+ b_{IJ}) \partial_{\pp} y^I \partial_=y^J~,
\eea
where the partial derivatives $\partial_=, \partial_{\pp}$ have been
taken with respect to the world-volume light-cone coordinates
$\sigma^{=}, \sigma^{\pp}$ of the string and we have identified the
embedding map of the string with the spacetime coordinates. It is
well-known that this action is invariant under the conformal
transformations $\delta y^M=\a(\sigma^{\pp}) \partial_{\pp} y^M+
\b(\sigma^{=}) \partial_= y^M$, where $\a, \b$ are infinitesimal
parameters. It remains to fix this residual symmetry in the
light-cone gauge, see e.g.~\cite{green}. For this, observe that the
equation of motion for $u$ implies that
$\partial_{\pp}\partial_=v=0$, i.e.~$v$ is a free boson. Then, one
can choose as the light-cone gauge condition  $v=p_{\pp}
\sigma^{\pp}+p_= \sigma^{=}$, where $p_{\pp}, p_=$ are
constants\footnote{If the string does not wrap around a circle, one
takes $p_{\pp}=p_=={p\over2}$.}. The light-cone Lagrangian reads
\bea
L_{{\rm l.c.}}=(g_{IJ}+ b_{IJ}) \partial_{\pp} y^I \partial_=y^J +2p_{\pp} p_= V
+   p_{\pp} (n_I+b_I) \partial_= y^I+p_= (n_I-b_I) \partial_{\pp} y^I~,
\la{lcactone}
\eea
where $H=db$ and $b= b_i e^-\wedge e^i+{1\over2} b_{ij} e^i\wedge e^j$.
As usual, the $u$ component of the embedding is determined by the vanishing of the two-dimensional energy-momentum tensor
$T_{\pp\pp}=T_{==}=0$ in terms of $y^I$.

The light-cone gauge fixing of  strings propagating in the
background (\ref{timeback}) is somewhat different. After writing the
world-sheet metric in conformal gauge, the bosonic part of the
Lagrangian of a string propagating in the background
(\ref{timeback}) is
\bea
L=2 f^4 (\partial_{\pp} v+ m_I \partial_{\pp} y^I)(\partial_= u+ n_I \partial_= y^I)+(g_{IJ}+ b_{IJ}) \partial_{\pp} y^I \partial_=y^J~.
\eea
The equations of motion for $u,v$ imply that
\bea
\partial_=[ f^4(\partial_{\pp} v+ m_I \partial_{\pp} y^I)]=0~,~~~\partial_{\pp}[f^4 (\partial_= u+ n_I \partial_= y^I)]=0~.
\eea
Therefore, the theory has two chiral $U(1)$ currents $\hat J_{\pp}=f^4(\partial_{\pp} v+ m_I \partial_{\pp} y^I)$
and $\check J_==f^4 (\partial_= u+ n_I \partial_= y^I)$.
The light-cone gauge fixing condition is chosen as
\bea
 \hat J_{\pp}={p_{\pp}\over2}~,~~~\check J_=={p_=\over2}~,
 \eea
 where $p_{\pp}, p_=$ are constants.
The energy momentum conditions $T_{\pp\pp}=T_{==}=0$ imply that
 \bea
&& p_{\pp} (\partial_{\pp} u+ n_I \partial_{\pp} y^I)+\gamma_{IJ}
\partial_{\pp} y^I \partial_{\pp}y^J=0~,
\cr
&& p_=(\partial_= v+ m_I \partial_= y^I)+\gamma_{IJ} \partial_=y^I
\partial_=y^J=0~.
\eea
Therefore, the light-cone gauge fixing conditions together with $T_{\pp\pp}=T_{==}=0$ determine $u,v$ in terms of $y$.
Finally, the light-cone action of $y$ is
\bea
L_{{\rm l.c.}}=(g_{IJ}+ b_{IJ}) \partial_{\pp} y^I \partial_=y^J- {1\over2}p_= p_{\pp}  f^{-4}+p_= m_I \partial_{\pp} y^I+p_{\pp} n_I \partial_= y^I~.
\la{lcacttwo}
\eea
Observe the similarity of the light-cone actions (\ref{lcactone}) and (\ref{lcacttwo}). This similarity
is  due to the T-duality between these type II backgrounds.

To summarize, the conformal symmetry of the world-sheet action of
the string can be fixed in the light-cone gauge for all backgrounds
with $\hat N,\check N\ge1$ supersymmetry. This does not extend to
generic $N\ge2$ backgrounds with either $\hat N=0$ or $\check N=0$.
A similar argument to the one we have used above reveals that for
those backgrounds only part of the conformal symmetry of the string
world-sheet action can be fixed in the light-cone gauge, see
\cite{ugplgp}.

\subsection{Spacetime supersymmetries and world-sheet W-symmetries}

One of the questions that arises is the relation between the
symmetries of the string world-sheet action and the spacetime
supersymmetry of the background in which the string propagates.
First, let us focus on the relation between world-sheet and
spacetime supersymmetry. As we have seen the eight-dimensional
manifold $B$ associated with $\hat N=\check N=1$ supersymmetric
backgrounds has either $Spin(7)$, $SU(4)$ or $G_2$ geometry. At
first sight this may suggest that the world-sheet action could admit
(1,1) supersymmetry which might enhance in the $SU(4)$ case to at
least (2,1) or (1,2). Indeed before light-cone gauge fixing, it is
straightforward to write the world-volume string action  in terms of
(1,1) superfields, see e.g.~\cite{green} and references therein.
 Moreover, there is a $W$-type of conserved current
  for every $\hat\nabla$- or $\check\nabla$-parallel form constructed from the Killing spinor bilinears
 $\hat\alpha$ and $\check\alpha$ \cite{howegpw}. Choosing the (super)conformal gauge for the world-sheet geometry, the string Lagrangian
 can be written as
 \bea
L=(g+b)_{MN} D_+Y^M D_-Y^N
\eea
in terms of (1,1) superfields $Y$.
The  currents  are
 \bea
 \hat J=\hat\alpha_{M_1\dots M_k}\, D_+Y^{M_1}\dots D_+Y^{M_k}~,~~~\check J=\check\alpha_{M_1\dots M_k}\, D_-Y^{M_1}\dots D_-Y^{M_k}~,
\eea
and  are conserved, $D_-\hat J=0$ and  $D_+\hat J=0$, subject to field equations, where
$D_+$ and $D_-$ are superspace derivatives, and  $D_-^2=i\partial_=$ and $D_+^2=i\partial_\pp$.

 The bilinears $\alpha$ constructed from one $\hat\nabla$- and one $\check\nabla$-parallel
 spinor are not associated with conserved world-sheet currents even though, as we have seen, they are instrumental in  understanding
 the geometry of the supersymmetric supergravity backgrounds. Applying this to the backgrounds with either $SU(4)$- or
 $SU(4)\ltimes\bR^8$-invariant spinors, one concludes  that the world-sheet supersymmetry does not enhance to either (2,1) or (1,2).
 This is because there is not a conserved current associated with the additional supersymmetry. Therefore this
 geometry is different from that found in the context of two-dimensional sigma models in \cite{wit}.

 After light-cone gauge fixing, the world-sheet supersymmetry
 of the light-cone action is not apparent. For example, the world-sheet light-cone action for the pp-wave type of backgrounds
 for which $\partial/\partial v$ is not Killing has an explicit dependence on the world-volume coordinates $\sigma^{=}, \sigma^{\pp}$.
 Therefore it cannot be supersymmetric. In addition, the light-cone actions
(\ref{lcactone}) and (\ref{lcacttwo}) have  scalar potential terms.
 To see whether these are compatible with world-sheet supersymmetry, first observe that
 $B$ does not necessarily admit a Killing vector field and then compare the scalar potential terms with those of (1,1)-supersymmetric
 sigma models with torsion found in \cite{towngp}. Compatibility would require that there was a function $h$ such that $|dh|^2=U$, where $U=-2 p_{\pp}p_= V$
 or $U={1\over2} p_{\pp}p_= f^{-4}$. So supersymmetrization of the world-sheet action depends on the existence of $h$.

\newsection{Concluding remarks}

We have determined the geometry of all type II common sector $N=2$
backgrounds. In particular, we have shown that the stability
subgroups of the Killing spinors in $Spin(9,1)$ for the type IIA
backgrounds are $SU(4)\ltimes\bR^8$ ($\hat N=2, \check N=0$), $G_2$
($\hat N=2, \check N=0$), $Spin(7)$ ($\hat N=1, \check N=1$),
$SU(4)$ ($\hat N=1, \check N=1$) and $G_2\ltimes\bR^8$ ($\hat N=1,
\check N=1$), where $\hat N$ and $\check N$ denote the positive and
negative chirality Killing spinors, respectively. The backgrounds
with $\check N=0$ are embeddings of $N=2$ backgrounds of the
heterotic string and their geometries have been analyzed in
\cite{ugplgp}. The spacetime of backgrounds with $Spin(7)$- and
$SU(4)$-invariant Killing spinors is a fibre bundle with fibre
directions given by the orbits of two null commuting Killing vectors
and with base space $B$ an eight-dimensional manifold with a
$Spin(7)$- and an $SU(4)$-structure, respectively. In particular, in
the $SU(4)$ case the almost complex structure of $B$ is not
integrable and the the fluxes depend on the trivialization of the
canonical bundle. The spacetime of backgrounds with
$G_2\ltimes\bR^8$-invariant Killing spinors is a pp-wave propagating
in an eight-dimensional manifold with a $G_2$-structure.
Alternatively, the spacetime can be viewed as a Lorentzian
deformation family of an eight-dimensional manifold with a
$G_2$-structure.

Similarly, we have shown that the stability subgroups of the Killing
spinors in $Spin(9,1)$ of the type IIB backgrounds are
$SU(4)\ltimes\bR^8$ ($\hat N=2, \check N=0$), $G_2$ ($\hat N=2,
\check N=0$), $Spin(7)\ltimes \bR^8$ ($\hat N=1, \check N=1$),
$SU(4)\ltimes \bR^8$ ($\hat N=1, \check N=1$) and $G_2$ ($\hat N=1,
\check N=1$). As in the IIA case, the backgrounds with $\check N=0$
are embeddings of $N=2$ backgrounds of the heterotic string and
their geometries have been analyzed in \cite{ugplgp}. The spacetime
of backgrounds with $Spin(7)\ltimes\bR^8$- and
$SU(4)\ltimes\bR^8$-invariant Killing spinors is a pp-wave
propagating in a $Spin(7)$ manifold and in an almost hermitian
manifold which admits an $SU(4)$-structure, respectively. The
spacetime of backgrounds with $G_2$-invariant spinors is a fibre
bundle with fibre given by the orbits of two commuting null Killing
vectors and with base space an eight-dimensional manifold with a
$G_2$-structure.

We have shown that there is a correspondence between type IIA and
IIB common sector backgrounds with $\hat N=\check N=1$
supersymmetry. In particular, to every type IIA  background with a
spacetime geometry that has an interpretation as a  Lorentzian
family of an eight-dimensional manifold  with a $K$-structure there
corresponds
 a type IIB background with spacetime geometry
that of   a rank two fibre bundle, and vice versa.
The geometry of the deformed manifold and that of the base space of the principal bundle
are the {\it same}. The structure groups of the spacetimes are interchanged as
\bea
K\leftrightarrow K\ltimes\bR^8
\eea
under this correspondence, where $K=Spin(7), SU(4)$ and $G_2$. This
correspondence may have been expected because of the type II
T-duality \cite{buscher, dai, dine, eric}. It is known that the
T-dual background of a fibre bundle along a spacelike fibre isometry
direction  is a trivial Lorentzian family, i.e.~it is  a pp-wave
background with two commuting isometries generated by the vector
fields $\partial/\partial v$ and $\partial/\partial u$. As we have
seen the correspondence
 persists after localization in the $v$ coordinate for the Lorentzian family because it does not change the
geometry of the deformed eight-dimensional manifold.

The question arises whether the geometries of common sector
backgrounds with $N>2$ supersymmetries can be classified as we have
done here for the $N=2$ backgrounds. The  investigation of
backgrounds with either $\hat N=0$ or $\check N=0$  reduces to that
of the heterotic string and therefore the result can be found in
\cite{ugplgp}. A preliminary investigation for the remaining cases,
$\hat N, \check N>0$ and $N=\hat N+\check N>2$, has revealed that
 more than one hundred new geometries can occur.
This is because there are many ways to embed the stability subgroups
$\hat G$ and $\check G$ of the Killing spinors in $Spin(9,1)$. This
leads to many different stability  subgroups $G=\hat G\cap \check G$
of all Killing spinors in $Spin(9,1)$ for the same number of
supersymmetries $N$. Nevertheless, all these geometries can be
classified using the techniques we have employed in this paper and
in \cite{ugplgp}. The completion of the programme will give an
understanding of the geometries of type II common sector backgrounds
with any number of supersymmetries.

\vskip 0.5cm
{\bf Acknowledgments:} We thank Jan Gutowski and Diederik Roest for many helpful discussions.
U.G. has a postdoctoral fellowship funded by
the Research Foundation K.U.~Leuven.

\setcounter{section}{0}

\appendix{Common sector field and Killing spinor equations}

The Killing spinor equations of the common sector of type II supergravities in components are
\bea
&&\hat\nabla\hat\epsilon=0~,~~~(\Gamma^M\partial_M\Phi-{1\over12}\Gamma^{MNP} H_{MNP})\hat\epsilon=0~,
\cr
&&\check\nabla\check\epsilon=0~,~~~(\Gamma^M\partial_M\Phi+{1\over12}\Gamma^{MNP} H_{MNP})\check\epsilon=0~,
\eea
where
\be
\hat\nabla_N Y^M=\nabla_N Y^M+{1\over2} H^M{}_{NR} Y^R~,
\ee
  $\nabla_M\epsilon=\partial_M\epsilon+{1\over4} \Omega_{M,AB} \Gamma^{AB}\epsilon$, and similarly for $\check \nabla$ after setting
  $H\rightarrow -H$,
and $\hat\epsilon$ and $\check\epsilon $ are Majorana-Weyl spinors
of the same (IIB) or opposite (IIA) chiralities. The integrability
conditions of the Killing spinor equations are
\bea
-2 E_{MN} \Gamma^N \hat\epsilon- e^{2\Phi} LH_{MN} \Gamma^N \hat\epsilon&=&0~,
\cr
L\Phi\hat\epsilon -{1\over4} e^{2\Phi} LH_{MN}\Gamma^{MN}\hat\epsilon&=&0~,
\cr
-2 E_{MN} \Gamma^N \check\epsilon+ e^{2\Phi} LH_{MN} \Gamma^N \check\epsilon&=&0~,
\cr
L\Phi\check\epsilon +{1\over4} e^{2\Phi} LH_{MN}\Gamma^{MN}\check\epsilon&=&0~,
\la{intcon}
\eea
where the field equations in the string frame to lowest order in
$\alpha'$  are
\bea
E_{MN}=R_{MN}-{1\over4} H_{PQM} H^{PQ}{}_N+2\nabla_M\partial_N\Phi&=&0~,
\cr
LH_{PQ}=\nabla_M (e^{-2\Phi} H^M{}_{PQ})&=&0~,
\cr
L\Phi=\nabla^2\Phi-2g^{MN} \partial_M\Phi \partial_N \Phi+{1\over12} H_{MNR} H^{MNR}&=&0~.
\eea
These integrability conditions are easily derived from those
of the heterotic string, see \cite{dewit, ugplgp}. We have also imposed the Bianchi identity of $H$, $dH=0$.

\appendix{Linear system}

To construct the linear systems associated with the Killing spinor
equations, one has to evaluate the supercovariant derivative and the
algebraic Killing spinor equations on a basis in the space of
spinors, for details see \cite{grangpb}. For the type IIB common
sector, this calculation can be read off from that of the heterotic
string in \cite{ugplgp}. Similarly, for the type IIA common sector,
 the evaluation of the $\hat\nabla$ and $d\Phi-{1\over2}H$ Killing spinor equations is identical to that of the heterotic string
 \cite{ugplgp}.
It remains to evaluate $\check\nabla$ and $d\Phi+{1\over2}H$ on a
basis of the negative chirality spinors $\Delta^-_{{\bf 16}}$
spanned by  forms of odd degree. A basis in the space of these
spinors is $e_5, e_i, e_{ijk}, e_{ij5}$ and $e_{12345}$. The
construction of the spinor representations and the spinor
 conventions can be found in \cite{ugplgp}.
In particular, one finds:

\bea
 \qu\check\Om_{A,BC}\G^{BC}e_5&=&\frac{1}{2\root}\left(\check\Om_{A,+-}+\check\Om_{A,k}{}^k\right)\G^+1-\frac{1}{\root}\check\Om_{A,-\bar{k}}\G^{\bar{k}}1
\cr &&
                         +\frac{1}{4\root}\check\Om_{A,\bar{k}\bar{l}}\G^+\G^{\bar{k}\bar{l}}1~,
\eea
\bea
 \qu\check\Om_{A,BC}\G^{BC}e_{12345}&=&\frac{1}{2\root}\left(\check\Om_{A,+-}-\check\Om_{A,k}{}^k\right)\G^+ e_{1234} -\frac{1}{8\root}\check\Om_{A,kl}\e^{kl}{}_{\bar{m}\bar{n}}\G^+\G^{\bar{m}\bar{n}}1\cr
                         && -\frac{1}{12\root}\check\Om_{A,-k}\e^k{}_{\bar{l}\bar{m}\bar{n}}\G^{\bar{l}\bar{m}\bar{n}}1~,
\eea
\bea
 \qu\check\Om_{A,BC}\G^{BC}e_{\a}&=&\frac{1}{2\root}\left(-\check\Om_{A,+-}
 +\check\Om_{A,\bar{\a}\a}+\check\Om_{A,k}{}^k\right)\G^{\bar{\a}}1
  +\frac{1}{\root}\check\Om_{A,+\a}\G^+1
  \cr
 &+&\frac{1}{2\root}\check\Om_{A,+\bar{k}}\G^+\G^{\bar{k}}\G^{\bar{\a}}1
  +\frac{1}{4\root}\check\Om_{A,\bar{k}\bar{l}}\G^{\bar{\a}}\G^{\bar{k}\bar{l}}1+\frac{1}{\root}\check\Om_{A,\bar{k}\a}\G^{\bar{k}}1~,
\eea

\bea
 \qu\check\Om_{A,BC}\G^{BC}\G^\a e_{1234}&=& \check\Om_{A,+\bar{\a}}\G^+ e_{1234}-
 \qu\check\Om_{A,\bar{\a}k}\e^k{}_{\bar{\a}\bar{l}\bar{m}}\G^{\bar{\a}\bar{l}\bar{m}}1
 \cr
  &+&\frac{1}{24}\left(-\check\Om_{A,+-}-\check\Om_{A,\bar{\a}\a}-\check\Om_{A,l}{}^l\right)
                                    \e^\a{}_{\bar{k}\bar{l}\bar{m}}\G^{\bar{k}\bar{l}\bar{m}}1
  \cr
  &+&\qu\check\Om_{A,+k}\e^{\a k}{}_{\bar{l}\bar{m}}\G^+\G^{\bar{l}\bar{m}}1
 -\ha\check\Om_{A,kl}\e^{\a kl}{}_{\bar{m}}\G^{\bar{m}}1~,
\eea
\bea
 \qu\check\Om_{A,BC}\G^{BC}\G^+ e_{\a\b}&=& -\check\Om_{A,\a\b}\G^+1+\ha\check\Om_{A,\bar k\bar l}\epsilon^{\bar k\bar l}{}_{\a\b}\G^+e_{1234}
 \cr
 & +&\qu\left(\check\Om_{A,+-}-\check\Om_{A,\g}{}^{\g}+\check\Om_{A,k}{}^k\right)\G^+\G_{\a\b}1 -\ha\check\Om_{A,-\bar k}\G_{\a\b}\G^{\bar k}1
 \cr
 &-&2\check\Om_{A,-[\a}\G_{\b]}1-\check\Om_{A,\bar k[\a}\G^+\G_{\b]}\G^{\bar k}1~,
\eea

\bea
 \left(d\Phi+\frac{1}{2}H\right)e_5&=& \root\left(\del_-\Phi+\ha H_{-k}{}^k\right)1
 -\frac{1}{\root}\left(\del_{\bar{k}}\Phi+\ha H_{+-\bar{k}}+\ha H_{\bar{k}l}{}^l\right)\G^+\G^{\bar{k}}1\cr
                 &+&\frac{1}{2\root}H_{-\bar{k}\bar{l}}\G^{\bar{k}\bar{l}}1-
                 \frac{1}{12\root}H_{\bar{k}\bar{l}\bar{m}}\G^+\G^{\bar{k}\bar{l}\bar{m}}1~,
\eea
\bea
 \left(d\Phi+\frac{1}{2}H\right)e_{12345}&=& \root\left(\del_-\Phi-\ha H_{-k}{}^k\right)e_{1234}
 \cr
 &-&\frac{1}{12\root}\left(\del_k\Phi+\ha H_{+-k}-\ha H_{kl}{}^l\right)\e^k{}_{\bar{m}\bar{n}\bar{p}}\G^+\G^{\bar{m}\bar{n}\bar{p}}1
 \cr
 &-&\frac{1}{4\root}H_{-kl}\e^{kl}{}_{\bar{m}\bar{n}}\G^{\bar{m}\bar{n}}1
 +\frac{1}{6\root}H_{klm}\e^{klm}{}_{\bar{n}}\G^+\G^{\bar{n}}1~,
\eea
\bea
 \left(d\Phi+\frac{1}{2}H\right)e_{\a}&=& \root\left(\del_\a\Phi-\ha H_{+-\a}+\ha H_{\a k}{}^k\right)1
 \cr
 & -&\frac{1}{\root}H_{+\a\bar{k}}\G^+\G^{\bar{k}}1 -
 \frac{1}{3\root}H_{\bar{k}\bar{l}\bar{m}}\e^{\bar{\a}\bar{k}\bar{l}\bar{m}}e_{1234}
 \cr
 &+&\frac{1}{\root}\left(\del_+\Phi+\ha H_{+\bar{\a}\a}+\ha H_{+k}{}^k\right)\G^+\G^{\bar{\a}}1
 \cr
 &+&\frac{1}{\root}\left(-\del_{\bar{k}}\Phi+\ha H_{+-\bar{k}}+\ha H_{\a\bar{\a}\bar{k}}-
 \ha H_{\bar{k}l}{}^l\right)\G^{\bar{\a}}\G^{\bar{k}}1
 \cr
 &+&\frac{1}{4\root}H_{+\bar{k}\bar{l}}\G^+\G^{\bar{\a}}\G^{\bar{k}\bar{l}}1+\frac{1}{2\root}H_{\a\bar{k}\bar{l}}\G^{\bar{k}\bar{l}}1~,
\eea
\bea
 \left(d\Phi+\frac{1}{2}H\right)\G^\a e_{1234}&=& 2\left(\del_{\bar{\a}}\Phi-\ha H_{\bar{\a}k}{}^k-\ha H_{+-\bar{\a}}\right)e_{1234}
 \cr
 & +&\frac{1}{12}\left(\del_+\Phi-\ha H_{+\bar{\a}\a}
 -\ha H_{+k}{}^k\right)\e^\a{}_{\bar{l}\bar{m}\bar{n}}\G^+\G^{\bar{l}\bar{m}\bar{n}}1
 \cr
 &+&\ha\left(\del_k\Phi-\ha H_{+-k}-\ha H_{\bar{\a}\a k}-\ha H_{kl}{}^l\right)\e^{\a k}{}_{\bar{l}\bar{m}}\G^{\bar{l}\bar{m}}1
 \cr
 & -&\ha H_{+kl}\e^{\a kl}{}_{\bar{m}}\G^+\G^{\bar{m}}1
 +\qu H_{+\bar{\a}k}\e^{\a k}{}_{\bar{l}\bar{m}}\G^+\G^{\bar{\a}}\G^{\bar{l}\bar{m}}1
 \cr
 &-&\frac{1}{3}H_{klm}\e^{\a klm}1-\ha H_{\bar{\a}kl}\e^{\a kl}{}_{\bar{m}}\G^{\bar{\a}}\G^{\bar{m}}1~,
\eea
\bea
\left(d\Phi+\frac{1}{2}H\right)\G^+ e_{\a\b}&=& -2H_{-\a\b}+H_{-\bar
k\bar l}\epsilon_{\a\b}{}^{\bar k\bar l}e_{1234}
\cr
 &+&\left(\del_-\Phi-\ha H_{-\g}{}^\g+\ha H_{-k}{}^k\right)\G_{\a\b}1
 \cr
 &+&2\left(-\del_{[\a}\Phi-\ha H_{+-[\a}+\ha H_{\g}{}^\g{}_{[\a}-\ha H_{
k}{}^k{}_{[\a}\right)\G^+\G_{\b]}1
 \cr
 &+&\ha\left(-\del_{\bar k}\Phi-\ha H_{+-\bar k}+\ha H_{\bar k\g}{}^\g-
 \ha H_{\bar k l}{}^{l}\right)\G^+\G_{\a\b}\G^{\bar k}1
 \cr
 &+&H_{\bar k\a\b}\G^+\G^{\bar k}1-2H_{-\bar k[\a}\G_{\b]}\G^{\bar k}1-\ha H_{\bar k\bar l [\a}\G^+\G_{\b]}\G^{\bar k\bar l}1~,
\eea
where   $A,B,C=\{+,-,\a,\bar\a,k,\bar k\}$, the range of the Greek
indices determined by the number of Greek indices on the left hand
side of the respective expression and the range of the Latin indices
given by ${\rm Range(\a)\cup {\rm Range(k)}=\{1,2,3,4\}}$. In
addition $d\Phi+{1\over2} H=\Gamma^A\partial_A\phi+{1\over12}
H_{ABC} \Gamma^{ABC}$.

\appendix{$SU(4)$ and $G_2$ geometries in eight dimensions}

\subsection{$SU(4)$-structures in eight dimensions}

$SU(4)$ geometries on an eight-dimensional manifold are
characterized by the existence of an almost complex structure $I$
compatible with a Riemannian metric $g$ and a $(4,0)$-form $\chi$
such that
\bea
{1\over 4!}\omega\wedge\omega\wedge\omega\wedge\omega= {1\over 2^{4}}\,\chi\wedge \bar \chi= d{\rm vol}~,~~~\omega\wedge \chi=0~,
\eea
where $\omega$ is the Hermitian, or K\"ahler, two-form. The
intrinsic torsion of such a manifold decomposes in terms of five
irreducible $SU(4)$ representations and therefore there are $2^{5}$
$SU(4)$-structures in an eight-dimensional manifold, see also
\cite{salamonb, cabrera, ugplgp}.
 These representations
can be found in the decomposition of $\nabla\omega$ and $\nabla\chi$ under $SU(4)$ representations, where $\nabla$ is the Levi-Civita
connection.
In particular,  one schematically has
\bea
\nabla_\a \omega_{\b\g}+{\rm c.c.}&\Longleftrightarrow &W_1+W_2
\cr
\nabla_{\bar\a}\omega_{\b\g}+\rm {c.c.}&\Longleftrightarrow &W_3+W_4
\cr
\nabla_{\bar\a} \chi_{\b_1\b_2\b_3\b_4}+\rm {c.c.}&\Longleftrightarrow& W_5
\eea
where $W_1$, $W_2$, $W_3$, $W_4$ and $W_5$ have dimensions 4, 20, 20, 4 and 4, respectively. These classes are also contained
in $d\omega$ and $d\chi$ as
\bea
d\omega^{3,0}+{\rm c.c.}&\Longleftrightarrow &W_1
\cr
d\chi^{3,2}+{\rm c.c.}&\Longleftrightarrow &W_1+W_2
\cr
d\omega^{2,1}+{\rm c.c.}&\Longleftrightarrow &W_3+W_4
\cr
d\chi^{4,1}+{\rm c.c.}&\Longleftrightarrow& W_4+W_5~.
\eea
The class $W_1$ is chosen such that $W_1=d\omega^{3,0}+d\omega^{0,3}$. This class can also be represented
by a one-form but this is special to the $SU(4)$ structure group.
Using this definition for $W_1$, one can write
\bea
\nabla_\a\omega_{\b\g}={1\over 3} (W_1)_{\a\b\g}+ (W_2)_{\a\b\g}~.
\eea
This equation can then be considered as the definition of $W_2$. The
class $W_4$ can be represented by the Lee one-form $\theta_\omega$
of $\omega$, i.e.
\bea
&&W_4=\theta_\omega=-\star(\star d\omega\wedge\omega)~.
\eea
We follow the form conventions of \cite{ugplgp}.
In turn  $W_3$ is defined by the relation
\bea
d\omega^{2,1}+d\omega^{1,2}= W_3+{1\over3} \omega\wedge W_4~.
\eea
Next consider the Lee form of ${\rm Re}\chi$
\bea
\theta_{{\rm Re}\chi}=-{1\over4}\star(\star d{\rm Re}\chi\wedge {\rm Re}\chi)~.
\eea
The Lee form  $\theta_{{\rm Re}\chi}$ can be decomposed in terms of the $W_4$ and $W_5$ representations. Since we have given a representative of
the $W_4$ representation,
we shall set
\bea
W_5=\theta_{{\rm Re}\chi}~.
\eea

Let us consider the change of the classes under conformal transformations $ds^2\rightarrow e^{2 f} ds^2$ of the metric and
changes in the trivialization of the canonical bundle ${\cal K}$. The latter are equivalent to
transforming $\chi\rightarrow e^{i\l} \chi$.
In particular one finds that
\bea
&&W_1^{f,\l}= e^{2 f} W_1~,~~~W_2^{f,\l}= e^{2 f} W_2~,~~~
W^{f,\l}_3=e^{2 f} W_3~,
\cr
&&W^{f,\l}_4=W_4+6df~,~~~W^{f,\l}_5=W_5-4d f-d_I\l~.
\eea
The only class that depends on the  trivialization of the canonical bundle is $W_5$.

\subsection{$G_2$-structures in eight dimensions}

The $G_2$ geometry of eight-dimensional manifolds is characterized by a $G_2$-invariant  three-form $\varphi$ and
a $G_2$ invariant non-vanishing one-form $Z$, such that
\bea
g(Z,Z)=1~,~~~{}^*\varphi\wedge Z=0~,~~~\varphi\wedge {}^*\varphi=-7\, d{\rm vol}~.
\eea
In particular ${}^*\varphi=Z\wedge \star\varphi$, where $\star\varphi$ is the standard $G_2$ invariant
four-form. The tangent bundle of the eight-dimensional manifolds decomposes as $TM=\bR\oplus E$, where $E$
is a rank 7 vector bundle. One can choose an orthonormal basis such that the metric on $M$ can be written
as
\bea
ds^2=(e^1)^2+\delta_{ij} e^i e^j~,~~~Z= e_1~,~~~i,j=2,\dots,8~,
\eea
To find the different $G_2$-structures in an eight-dimensional manifold, one can use
the method proposed in \cite{grayhervella}.
The intrinsic torsion of such a manifold decomposes in terms of ten irreducible $G_2$
representations. So there are $2^{10}$ $G_2$-structures in an eight-dimensional manifold. These representations
can be found by decomposing  $\nabla Z$ and $\nabla\varphi$ in terms of $G_2$ representations. In particular, one has that
\bea
\nabla_1 Z_i&\Longleftrightarrow &X
\cr
\nabla_1\varphi_{ijk}&\Longleftrightarrow &W
\cr
\nabla_i Z_j&\Longleftrightarrow& X_1+X_2+X_3+X_4
\cr
\nabla_i \varphi_{jkl}&\Longleftrightarrow &W_1+W_2+W_3+W_4
\eea
where $X$ and $W$ both have dimension 7, $X_1$, $X_2$, $X_3$ and
$X_4$ have dimensions 1, 7, 14 and 27, respectively, and similarly
for $W_1, W_2, W_3$ and $W_4$. It is straightforward to find $X, W$,
$X_1$, $X_2$, $X_3$, $X_4$, $W_1, W_2, W_3$ and $W_4$ in terms of
$\nabla Z$ and $\nabla\varphi$.


\begin{thebibliography}{00}
\addcontentsline{toc}{section}{References} \frenchspacing \small
\addtolength{\itemsep}{-4pt}

\bibitem{ugplgp}
  U.~Gran, P.~Lohrmann and G.~Papadopoulos,
  ``The spinorial geometry of supersymmetric heterotic string backgrounds,''
  arXiv:hep-th/0510176.

\bibitem{grangp}
J.~Gillard, U.~Gran and G.~Papadopoulos,
``The spinorial geometry of supersymmetric backgrounds,''
Class.\ Quant.\ Grav.\  {\bf 22} (2005) 1033
[arXiv:hep-th/0410155].



\bibitem{strominger}
A.~Strominger, ``Superstrings with torsion,'' Nucl.\ Phys.\ B {\bf
274}, 253 (1986).

\bibitem{hullb}
C.~M.~Hull, ``Compactifications of the heterotic superstring,''
Phys.\ Lett.\ B {\bf 178} (1986) 357.


\bibitem{tseytlin}
G.~Papadopoulos and A.~A.~Tseytlin,
``Complex geometry of conifolds and 5-brane wrapped on 2-sphere,''
Class.\ Quant.\ Grav.\  {\bf 18} (2001) 1333
[arXiv:hep-th/0012034].




\bibitem{callan}
C.~G.~.~Callan, J.~A.~Harvey and A.~Strominger,
``Supersymmetric string solitons,''
arXiv:hep-th/9112030.



\bibitem{tsimpis}
J.~Gillard, G.~Papadopoulos and D.~Tsimpis,
``Anomaly, fluxes and (2,0) heterotic-string compactifications,''
JHEP {\bf 0306} (2003) 035
[arXiv:hep-th/0304126].

\bibitem{lust}
G.~L.~Cardoso, G.~Curio, G.~Dall'Agata and D.~Lust,
``BPS action and superpotential for heterotic string compactifications  with
fluxes,''
JHEP {\bf 0310} (2003) 004
[arXiv:hep-th/0306088].


\bibitem{howegpc}
P.~S.~Howe and G.~Papadopoulos,
``Twistor spaces for HKT manifolds,''
Phys.\ Lett.\ B {\bf 379} (1996) 80
[arXiv:hep-th/9602108].

P.~S.~Howe, A.~Opfermann and G.~Papadopoulos,
``Twistor spaces for QKT manifolds,''
Commun.\ Math.\ Phys.\  {\bf 197} (1998) 713
[arXiv:hep-th/9710072].

\bi{poon} G.~ Grantcharov and  Y.~ S.~ Poon, ``Geometry of
hyper-K\"ahler connections with torsion'', Commun.Math.Phys. {\bf
213} (2000) 19-37, [math.DG/9908015].

\bibitem{ivanovgp}
S.~Ivanov and G.~Papadopoulos,
``A no-go theorem for string warped compactifications,''
Phys.\ Lett.\ B {\bf 497} (2001) 309
[arXiv:hep-th/0008232].

``Vanishing theorems and string backgrounds,''
Class.\ Quant.\ Grav.\  {\bf 18} (2001) 1089
[arXiv:math.dg/0010038].

\bibitem{ivanovgpb}
J.~Gutowski, S.~Ivanov and G.~Papadopoulos,
``Deformations of generalized calibrations and compact non-Kahler manifolds
with vanishing first Chern class,'' Asian J. Math. {\bf 7} (2003), 39-80
[arXiv:math.dg/0205012].

\bi{salamon}  A.~ Fino, M.~ Parton, S.~ Salamon, ``Families of strong KT structures in six dimensions'',
[math.DG/0209259].


\bibitem{goldstein}
E.~Goldstein and S.~Prokushkin,
``Geometric model for complex non-Kaehler manifolds with SU(3) structure,''
Commun.\ Math.\ Phys.\  {\bf 251} (2004), 65
[arXiv:hep-th/0212307].

\bibitem{poonb}
D.~Grantcharov, G.~Grantcharov, Y.~S.~Poon, ``Calabi-Yau connections
with torsion on toric bundles'' [math.DG/0306207].



 \bi{stefang2} T. Friedrich, S. Ivanov, ``Parallel spinors and connections with skew-symmetric torsion in string theory''
 Asian Journal of Mathematics {\bf  6} (2002), 303-336 [math.DG/0102142].

``Killing spinor equations in dimension 7 and geometry of integrable
$G_2$-manifolds''
 J. Geom. Phys. 48 (2003), 1-11 [math.DG/0112201].

\bi{stefanspin7} S.~ Ivanov, ``Connection with torsion, parallel
spinors and geometry of $Spin(7)$ manifolds''
 Math. Res. Lett. {\bf 11} (2004), no. 2-3, 171--186 [math.DG/0111216].



\bibitem{gauntlett}
J.~P.~Gauntlett, D.~Martelli, S.~Pakis and D.~Waldram,
``G-structures and wrapped NS5-branes,''
Commun.\ Math.\ Phys.\  {\bf 247} (2004) 421
[arXiv:hep-th/0205050].


J.~P.~Gauntlett, D.~Martelli and D.~Waldram,
``Superstrings with intrinsic torsion,''
Phys.\ Rev.\ D {\bf 69} (2004) 086002
[arXiv:hep-th/0302158].


\bibitem{lustb}
G.~L.~Cardoso, G.~Curio, G.~Dall'Agata, D.~Lust, P.~Manousselis and G.~Zoupanos,
``Non-Kaehler string backgrounds and their five torsion classes,''
Nucl.\ Phys.\ B {\bf 652} (2003) 5
[arXiv:hep-th/0211118].



\bi{jose} J.~M.~Figueroa-O'Farrill, ``Breaking the M-waves,''
Class.\ Quant.\ Grav.\  {\bf 17} (2000) 2925 [arXiv:hep-th/9904124].


\bibitem{grangpb}
U.~Gran, G.~Papadopoulos and D.~Roest,
``Systematics of M-theory spinorial geometry,''
Class.\ Quant.\ Grav.\  {\bf 22} (2005) 2701
[arXiv:hep-th/0503046].

U.~Gran, J.~Gutowski, G.~Papadopoulos and D.~Roest, ``Systematics of
IIB spinorial geometry,'' Class.\ Quant.\ Grav.\  {\bf 23} (2006)
1617 [arXiv:hep-th/0507087].

\bibitem{ruiz}
A.~Dabholkar, G.~W.~Gibbons, J.~A.~Harvey and F.~Ruiz Ruiz,
``Superstrings and solitons,'' Nucl.\ Phys.\ B {\bf 340} (1990) 33.

\bibitem{green}
  M.~B.~Green, J.~H.~Schwarz and E.~Witten,
  ``Superstring Theory. Vol. 1: Introduction,'' Cambridge Monographs On Mathematical Physics (1987).


\bibitem{howegpw}
  P.~S.~Howe and G.~Papadopoulos,
  ``Holonomy groups and W symmetries,''
  Commun.\ Math.\ Phys.\  {\bf 151} (1993) 467
  [arXiv:hep-th/9202036].


  \bibitem{wit}
  B.~de Wit and P.~van Nieuwenhuizen,
  ``Rigidly and locally supersymmetric two-dimensional nonlinear sigma models
  with torsion,''
  Nucl.\ Phys.\ B {\bf 312} (1989) 58.

   G.~W.~Delius, M.~Rocek, A.~Sevrin and P.~van Nieuwenhuizen,
  ``Supersymmetric sigma models with nonvanishing Nijenhuis tensor and their
  operator product expansion,''
  Nucl.\ Phys.\ B {\bf 324} (1989) 523.


\bibitem{towngp}
  C.~M.~Hull, G.~Papadopoulos and P.~K.~Townsend,
  ``Potentials for (p,0) and (1,1) supersymmetric sigma models with torsion,''
  Phys.\ Lett.\ B {\bf 316} (1993) 291
  [arXiv:hep-th/9307013].




\bibitem{buscher}
  T.~H.~Buscher,
  ``A symmetry of the string background field equations,''
  Phys.\ Lett.\ B {\bf 194} (1987) 59.

\bibitem{dai}
  J.~Dai, R.~G.~Leigh and J.~Polchinski,
  ``New connections between string theories,''
  Mod.\ Phys.\ Lett.\ A {\bf 4} (1989) 2073.

  \bibitem{dine}
  M.~Dine, P.~Y.~Huet and N.~Seiberg,
  ``Large and small radius in string theory,''
  Nucl.\ Phys.\ B {\bf 322} (1989) 301.

\bibitem{eric}
  E.~Bergshoeff, C.~M.~Hull and T.~Ortin,
  ``Duality in the type II superstring effective action,''
  Nucl.\ Phys.\ B {\bf 451} (1995) 547
  [arXiv:hep-th/9504081].


\bibitem{dewit}
B.~de Wit, D.~J.~Smit and N.~D.~Hari Dass, ``Residual supersymmetry
of compactified D = 10 supergravity,'' Nucl.\ Phys.\ B {\bf 283}
(1987) 165.


\bi{salamonb} S. Chiossi and S. Salamon, ``The intrinsic torsion of
$SU(3)$
 and $G_2$-structures'', Diff. Geom., Valencia 2001, World Sci. 2002, 115
 [arXiv:math.DG/0202282].

\bibitem{cabrera} F.~M.~ Cabrera, ``Special almost Hermitian geometry,''
math.DG/0409167



\bi{grayhervella}
A. Gray and L.M. Hervella, ``The sixteen classes of
 almost Hermitian manifolds and their linear invariants'',
 Ann.\ Mat.\ Pura\ e \ Appl.\ {\bf 282} (1980) 1.








\end{thebibliography}
\end{document}